\documentclass[aps,twocolumn,reprint,amsmath,amssymb,superscriptaddress,floatfix,prb]{revtex4-2}
\usepackage[breaklinks=true,colorlinks,citecolor=blue,linkcolor=blue,urlcolor=blue]{hyperref}
\usepackage{epsfig,mathrsfs,color,latexsym,subfigure,marginnote,graphicx,verbatim,relsize,mathrsfs,color,array,amsmath,amsfonts,amssymb,graphicx}

\begin{document}
	
\title{Coexistence of phononic Weyl, nodal line, and threefold excitations \\ in chalcopyrite CdGeAs$_{2}$ and associated thermoelectric properties}
%\title{Topological phonons and thermoelectric studies of a chalcopyrite system CdGeAs$_{2}$}

\author{Vikas Saini}
\email{vikas.saini@tifr.res.in}
\affiliation{Department of Condensed Matter Physics and Materials Science,Tata Institute of Fundamental Research, Mumbai 400005, India}
	
\author{Bikash Patra}
\email{bikash.patra@tifr.res.in}
\affiliation{Department of Condensed Matter Physics and Materials Science,Tata Institute of Fundamental Research, Mumbai 400005, India}
	
\author{Bahadur Singh}
%\email{bahadur.singh@tifr.res.in}
\affiliation{Department of Condensed Matter Physics and Materials Science,Tata Institute of Fundamental Research, Mumbai 400005, India}
	
\author{A. Thamizhavel}
\email{thamizh@tifr.res.in} 
\affiliation{Department of Condensed Matter Physics and Materials Science,Tata Institute of Fundamental Research, Mumbai 400005, India}

\begin{abstract}
Realization of topologically protected quantum states leads to unprecedented opportunities for fundamental science and device applications. Here, we demonstrate the coexistence of multiple topological phononic states and calculate the associated thermoelectric properties of a chalcopyrite material CdGeAs$_2$ using first-principles theoretical modeling. CdGeAs$_{2}$ is a direct bandgap semiconductor with a bandgap of $0.65$ eV. By analysing the phonon spectrum and associated symmetries, we show the presence of nearly isolated Weyl, nodal line, and threefold band crossings in CdGeAs$_2$. Specifically,  the two triply degenerate points (TDPs) identified on the $k_{z}$ axis are formed by the optical phonons bands 7, 8, and 9 with type-II energy dispersion. These TDPs form a time reversal pair and are connected by a straight nodal line with zero Berry phase. The TDPs formed between bands 14, 15, and 16 exhibit type-I crossings and are connected through the open straight nodal line. Our transport calculations show a large thermopower exceeding $\sim$500 and $200$ $\rm \mu V/K$ for the hole and electron carriers, respectively, above 500 K with a carrier doping of 10$^{18}$ cm$^{-3}$. The large thermopower in $p$-type CdGeAs$_{2}$ is a consequence of the sharp density of states appear from the presence of a heavy hole band at the $\Gamma$ point. We argue that the presence of topological states in the phonon bands could lead to low lattice thermal conductivity and drive a high figure-of-merit in CdGeAs$_{2}$.
\end{abstract}

\maketitle		

\section{Introduction} \label{sec:sample1}

The existence of topologically protected nontrivial states has been demonstrated in wide classes of crystalline materials~\cite{AMPER2022_BS, RMP_Bansil, RMP_Hasan}. For example, Dirac semimetals~\cite{liu2014discovery, jenkins2016three, nie2018phase, wang2012dirac} are formed by the crossings of spin degenerate valence and conduction bands in the momentum space. The low-energy excitations around the crossing points mimic the linear energy-momentum relation of Dirac fermions. These Dirac points are invariant under perturbations preserving parity and time-reversal symmetries. Breaking either of these symmetries splits a Dirac point into two Weyl points with opposite chirality~\cite{sun2015topological, yan2017topological, xu2015discovery, lv2015observation}. The band crossings in presence of non-symmorphic crystallographic symmetries can lead to higher fold fermionic excitations and have been realized in numerous materials~\cite{yang2020observation,thirupathaiah2021sixfold, yang2019topological,hasan2021weyl}. The triply degenerate point (TDP) semimetal states have been established in MoP, WC, ZrSe, etc. materials where the presence of three-fold crossings are shown to exist between a double degenerate and single degenerate bands~\cite{lv2017observation, kumar2019extremely, zhu2016triple, ma2018three, weng2016topological, Tripple_BaAgAs}. Importantly, the TDP semimetal phase conceptually lies between the Dirac and Weyl phases. Topological band crossings can also be classified based on the dimensionality of the band touching points.  They can form nodal points with zero-dimensional point-like crossings or nodal lines with one-dimensional band crossings. Such one-dimensional nodal lines can constitute nodal lines, nodal links, nodal knots, nodal rings, and nodal chains, etc.~\cite{yan2017nodal, bi2017nodal, fang2016topological, rui2018topological, zhu2022symmetry}. Besides the degeneracies and dimensionalities of the crossing points, chiral charges of the nodal points, energy dispersion and slope of the bands are essential to distinguish among a variety of topological states. In particular, the type-II band crossings are identified at the touching point of the hole and electron pockets and break the Lorentz symmetry~\cite{zhang2017type, soluyanov2017type, sun2015prediction, xia2019symmetry, LaAlGe_typeII}. The topology in the electronic structure has been explored over the last several years, and many novel phenomena have been proposed and verified in experiments. Common to the topology of electronic structure is that they are constrained by the Pauli exclusion principle. 

Topology of the bosonic states especially topological states in the phonon spectrum is an emerging research field. Phonons obey the Bose-Einstein statistics and are not constrained by the Fermi energy, giving access to the whole spectrum of phonon energy range from the THz and IR to probe the topological states~\cite{zhu2018observation, wang2021coexistence, litvinchuk2020raman}. Recent studies on graphene, FeSi, and other materials uncover the presence of topological excitations such as Weyl, double Weyl, and multifold Weyl in the phonon spectrum. The higher fold degenerate phonons have been proposed for several materials with specific space groups which show the topological phonon states protected by various crystalline symmetries~\cite{li2019topological, miao2018observation, liu2021charge, wang2021symmetry,xie2021sixfold, liu2020topological, zhong2021coexistence}. Motivated by the studies of topological phonons in materials and their possible effects on thermal transport, we explore the topological phonons and transport properties of chalcopyrite material CdGeAs$_{2}$. Our phonon calculations reveal topological Weyl phonons, triply-degenerate nodal points, and nodal lines, among other phases in various phonon bands. Specifically,  we find Weyl points with chiral charge $\pm 1$ in multiple phonons branches. Moreover, multiple TDPs protected by $C_{3v}$ symmetry are found on the $k_z$ axis with both the type I and type II energy dispersions. 

We also investigate the electronic and thermoelectric properties of CdGeAs$_{2}$. Electronic structure and transport calculations are carried with both the generalized gradient approximation (GGA)~\cite{perdew1996generalized} and mBJ~\cite{tran2009accurate} exchange-correlation (XC) functionals and with the inclusion of spin-orbit coupling (SOC). The results obtained with mBJ functional shows a band gap of 0.65 eV, close to experimentally observed values of 0.57 eV at room temperature~\cite{mccrae1997photoluminescence,akimchenko1973electroreflection, bai2005urbach}. The calculated thermopower ($S$) of $p$-type carriers is found to be more than the $n$-type carriers in our considered carrier density range of 10$^{18}$ - 10$^{21}$ cm$^{-3}$. This large thermopower for $p$-type carriers is attributed to the steepness in the density of states (DOS) below the Fermi energy level due to the presence of heavy hole-type bands. We also discuss the effect of carrier effective mass in generating the large thermopower and figure of merit $ZT{_e}$ $=$ $\frac{S^{2}\sigma/\tau}{k_{e}/\tau}$. Specifically, the upper limit of $ZT{_e}$ without considering the lattice thermal conductivity, is remarkably large with a value of $\sim11.5$ at 600 K and carrier density $\sim$ 1.1$\times$ 10$^{18}$ cm$^{-3}$ for $p$-type carriers. In this way, our work identifies CdGeAs${_2}$ as a potential chalcopyrite material for exploring topological phonons and thermoelectric properties. 

\begin{figure}[ht!]
\includegraphics[width=0.45\textwidth]{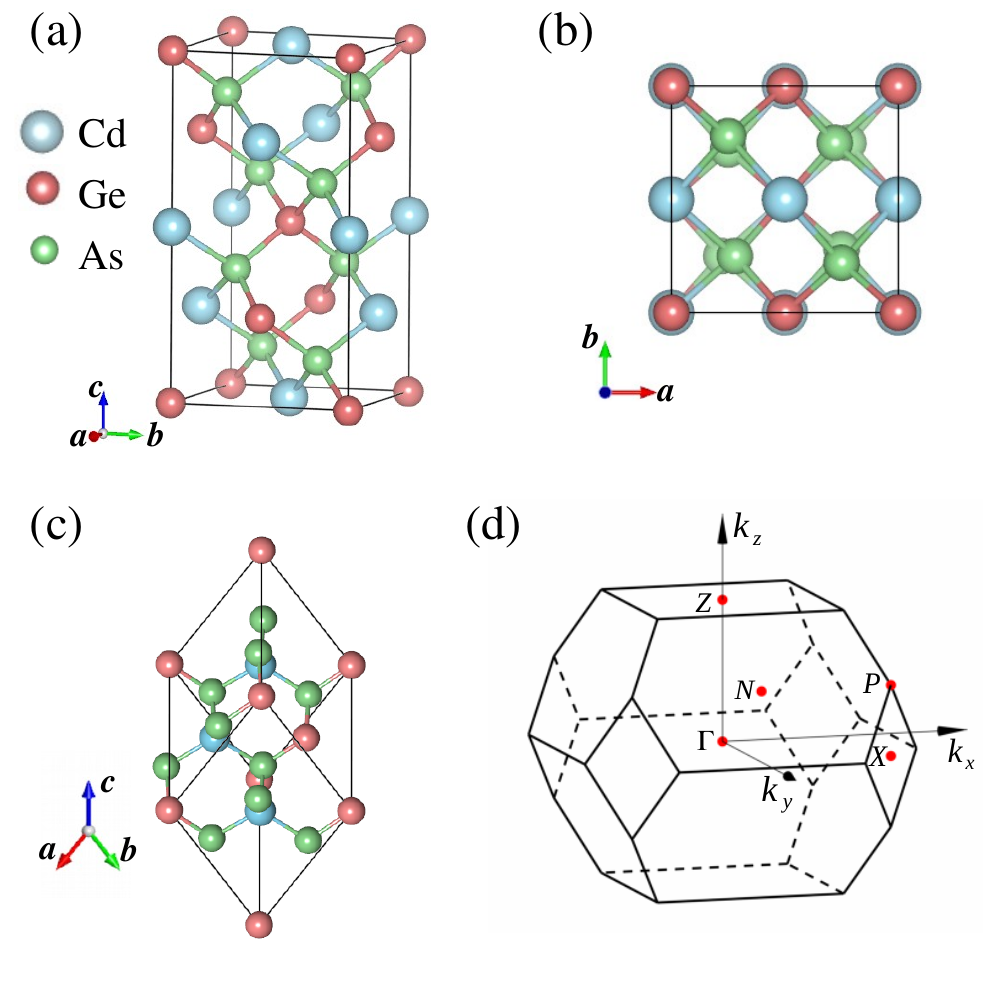}
\caption{Crystal structure and Brillouin zone of CdGeAs$_{2}$. (a), (b) Conventional tetragonal unit cell of CdGeAs$_{2}$. (c) Primitive unit cell and (d) associated Brillouin zone structure of CdGeAs$_{2}$. High-symmetry points are marked in red.}
\label{Fig1}
\end{figure}   
	
\section{Methods}

Electronic structure calculations were performed within the framework of density functional theory (DFT) using the full-potential linearized augmented plane wave (FP-LAPW) method as implemented in WIEN2k package~\cite{blaha2001wien2k,schwarz2003solid,schwarz2003dft}. Self-consistent calculations were performed on a dense $k$-mesh with $10000$ $k$ points. Plane-wave cut-off for $\rm R_{MT}K_{MAX}$ was set to be 9 (R$_{MT}$ is the muffin tin radius and K$_{MAX}$ is the maximum value of reciprocal lattice vector). Thermoelectric properties were calculated by solving the Boltzmann transport equations under the constant scattering time approximation (CSTA) as implemented in the BoltzTraP2~\cite{madsen2006boltztrap, madsen2018boltztrap2}.  We used $149784$ irreducible $k$ points to accurately model transport properties. The phonon spectrum was obtained using the finite displacement method by considering a 2$\times$2$\times$2 supercell employing the VASP~\cite{kresse1993ab, kresse1996efficient, kresse1999from} and Phonopy codes~\cite{togo2015first}. The lattice parameters for the first-principles and thermoelectric calculations were obtained from the powder x-ray diffraction experiments using our synthesized CdGeAs{$_2$} polycrystalline sample. The refined parameters were obtained as $a$ = 5.94(4) \r{A}, $c$ = 11.21(6) \r{A} and the Wyckoff positions (0.22009, 0.25, 0.125), (0.0, 0.0, 0.5), and (0.0, 0.0, 0.0) for As, Cd, and Ge atoms, respectively.  
	
\section{Results and Discussion}
\begin{figure*}%[]
\includegraphics[width=1.0\textwidth]{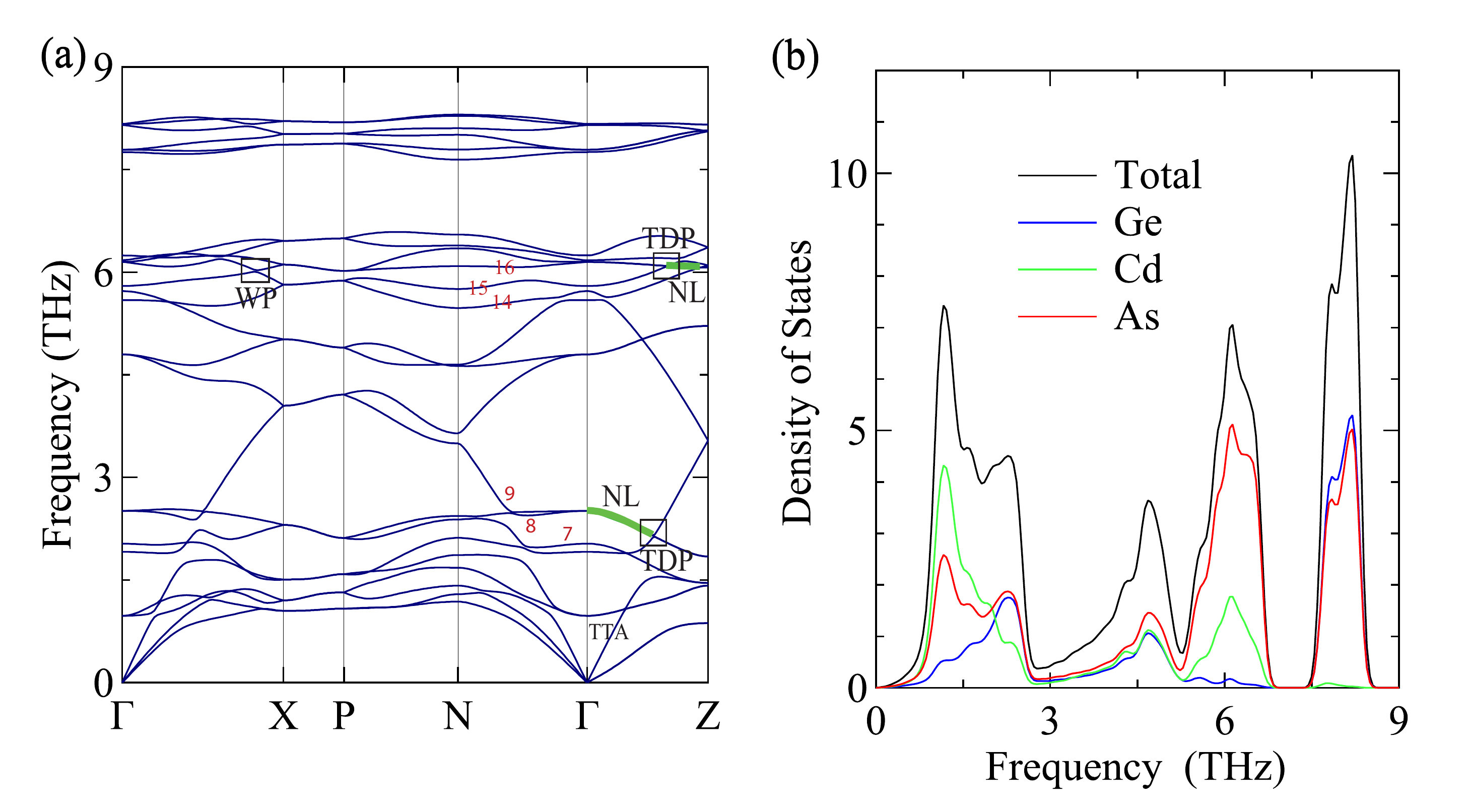}
\caption{(a) Calculated phonon dispersion of CdGeAs$_2$ along high symmetry Brillouin zone directions. The topological crossings are highlighted. (b) Partial and total phonon density of states (DOS) of CdGeAs$_{2}$.}\label{Fig2}
\end{figure*}

\begin{figure*}%[]
\includegraphics[width=0.9\textwidth]{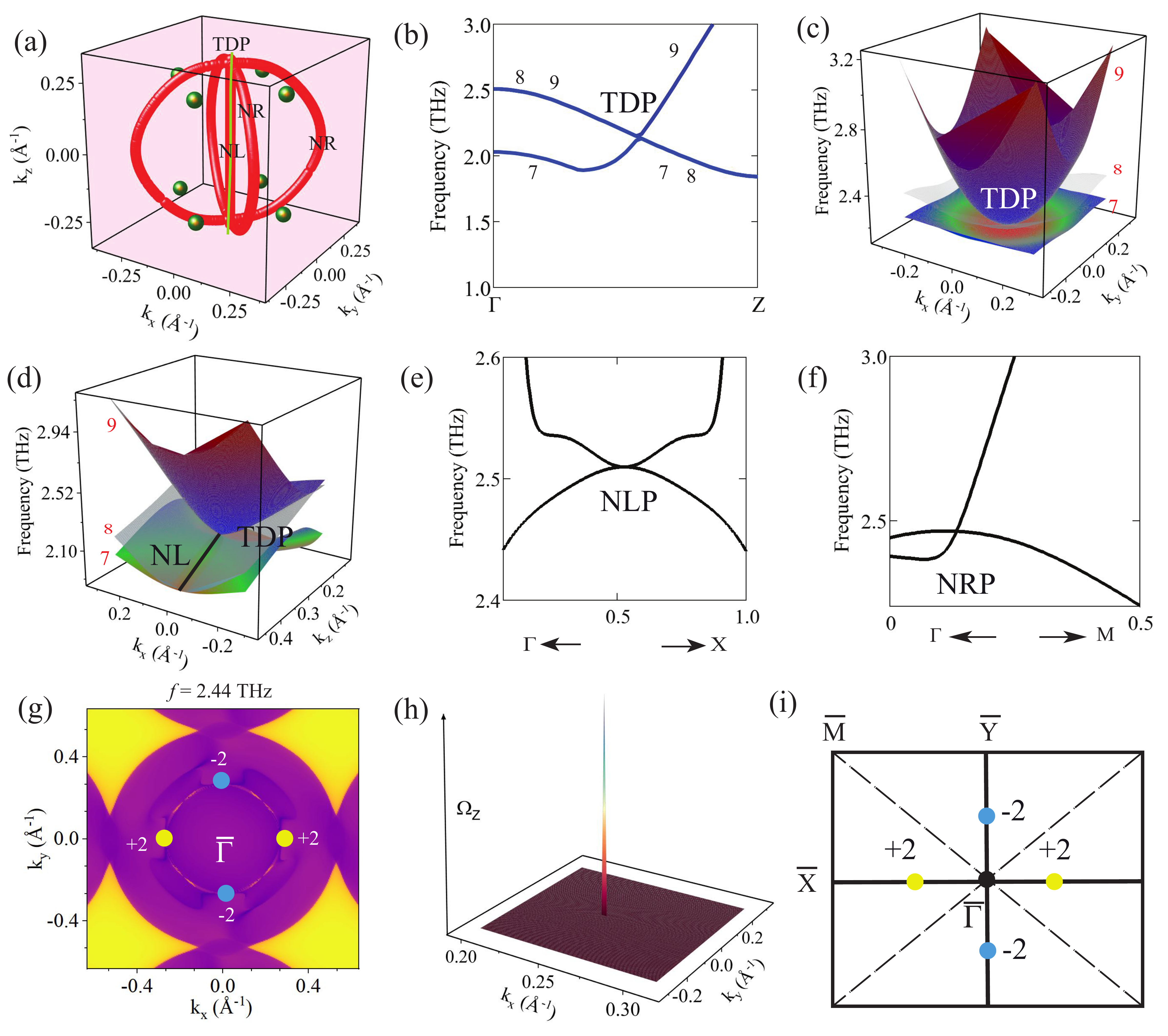}
\caption{Topological phonon structure for bands 8 and 9 (see Fig. ~\ref{Fig2} for band index). (a) Nodal crossing points formed by the bands $8$ and $9$ in 3D momentum space. (b) Triple degenerate point and nodal line along $\Gamma$-$Z$ direction. (c-d) 3D dispersion around the TDP in $k_{x}-k_{y}$ and $k_{x}-k_{z}$ planes. (e) Phonon dispersion for a nodal line point along in-plane axis. (f) Phonon dispersion for a point present at nodal ring along the diagonal axis. (g) Topological fermi arc surface states connecting the projected phonon Weyl points on the (001) surface. (h) Non-zero Berry curvature around the positive Weyl point. (i) Schematic diagram of the Weyl points on the (001) surface. Blue and yellow circles denote the positive and negative charges.}\label{Fig3}
\end{figure*}

\begin{figure*}%[!]
\includegraphics[width=0.9\textwidth]{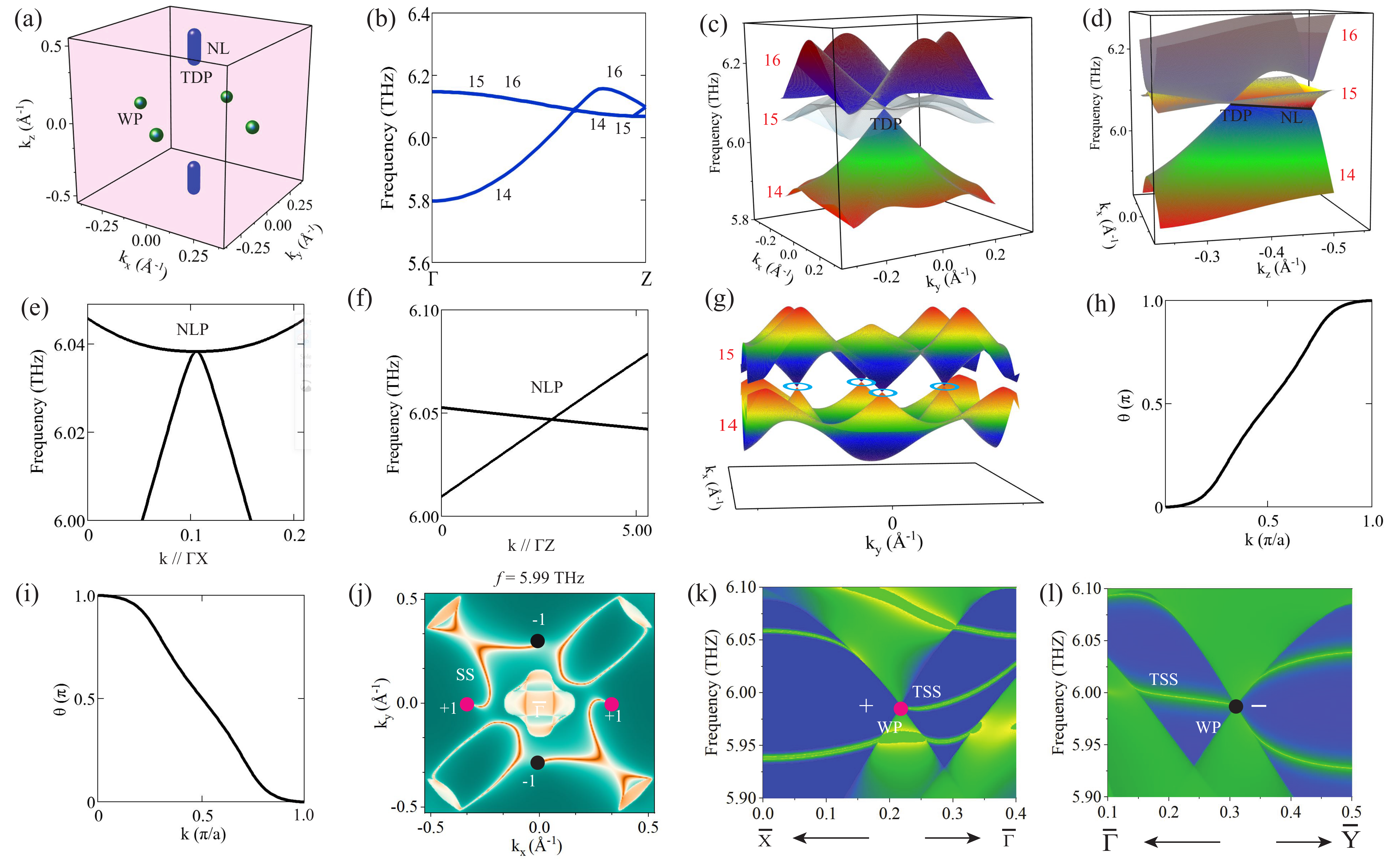}
\caption{Topological phonon structure for bands 14 and 15 (see Fig. ~\ref{Fig2} for band index). (a) Nodal crossings of bands 14 and 15 inside the BZ. (b) Isolated part for the TDP and NL along $\Gamma$-$Z$ direction. (c-d) 3D phonon dispersion in $k_{x}-k_{y}$ and $k_{x}-k_{z}$ planes around the TDP. NL is marked along the $k_{z}$ axis in (d). (e-f) Dispersion along the  $k_{y}$ and $k_{z}$ directions around a point present at the nodal line. (g) Weyl points distribution on the (001) surface. The WPs are drawn as circles. (h-i) Wannier center evolution on a sphere enclosing the positive and negative chiral charge WPs. (d) Topological fermi arc surface states connecting opposite chiral WPs on the (001) surface. (k-l) Chiral topological surface states connecting the projected bulk WPs. }\label{Fig4}	
\end{figure*}

\begin{figure*}%[!]
\includegraphics[width=1.0\textwidth]{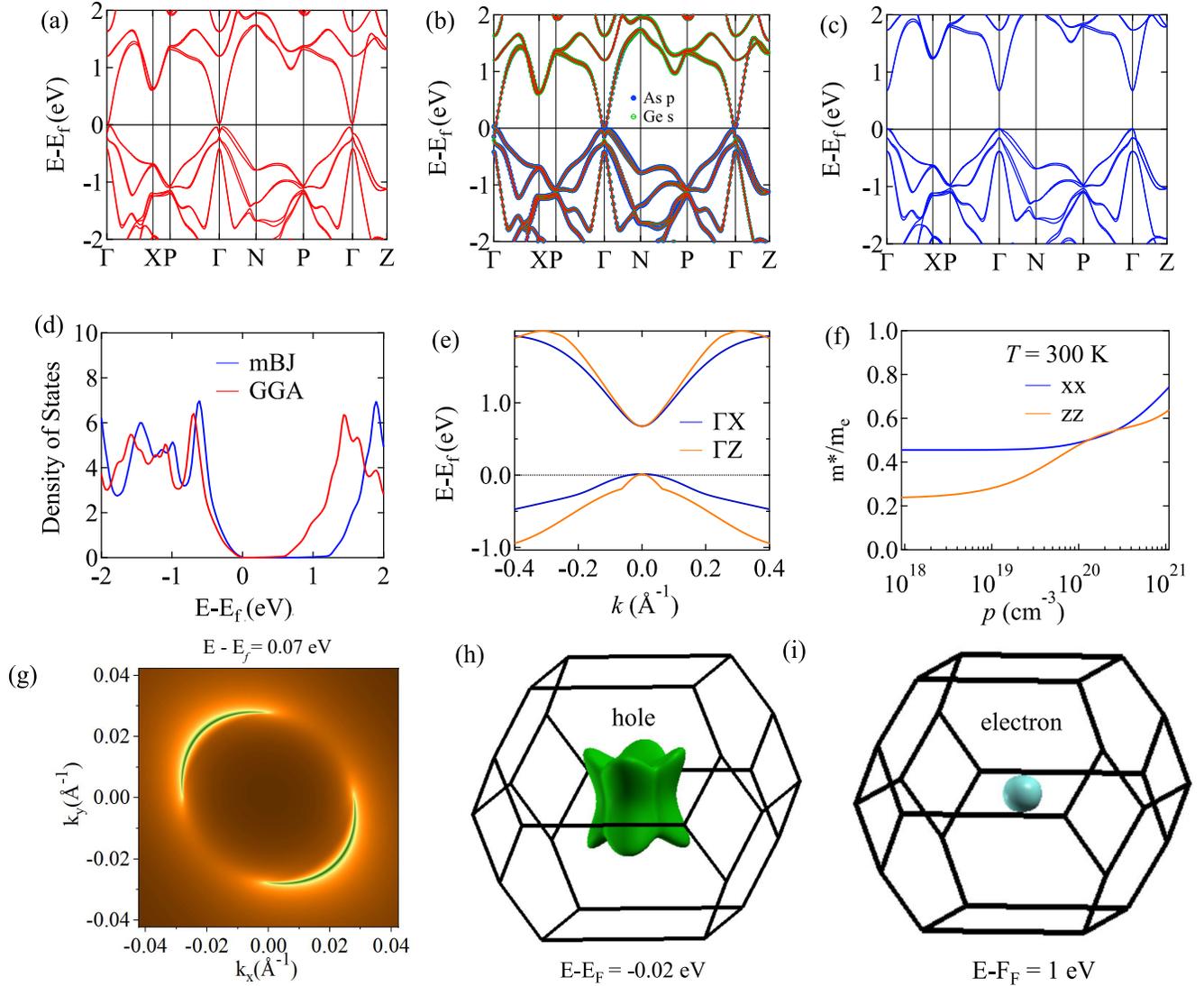}
\caption{(a) Electronic band structure of CdGeAs$_{2}$ obtained with GGA XC functional.  (b) Orbital resolved band structure obtained with GGA XC functional. (c) Same as (a) but obtained with the mBJ XC functional. (d) Calculated total DOS for the GGA and mBJ XC functionals. (e) Band dispersion of valence and conduction bands for $\Gamma$-$X$ and $\Gamma$-$Z$ directions. (f) Calculated carrier effective masses for in-plane and out-of-plane directions for $p$-type carriers at 300 K. (g) Topologically nontrivial surface states on crystallographic (001) surface obtained for GGA XC functional without SOC. (h) Anisotropic nature of topmost hole band below 20~meV energy from the Fermi level for mBJ XC functional calculation. (i) Electron pocket of the lowest conduction band at an energy around 1~eV upward from the Fermi level for mBJ XC functional.}
\label{Fig5}
\end{figure*}

The CdGeAs$_{2}$ crystallizes in a non-centrosymmetric tetragonal crystal lattice with space group $I$$\bar42d$ (No. 122). The conventional unit cell of CdGeAs$_{2}$ is shown in Fig.~\ref{Fig1}(a). The As atoms in the top layer are shifted with respect to the As atoms in the bottom layer, as shown in Fig.~\ref{Fig1}(b). This suggests the absence of inversion symmetry in the crystal. Figure~\ref{Fig1}(c) shows the primitive unit cell and Fig.~\ref{Fig1}(d) illustrates the associated Brillouin zone (BZ) with high-symmetry points marked. The phonon dispersion of CdGeAs$_{2}$ along various high symmetry paths is shown in Fig.~\ref{Fig2}(a). The absence of negative phonon frequencies suggests the dynamical stability of the crystal. From the phonon calculations, we found that the acoustic phonons are mainly contributed by the heavy Cd atoms, as depicted from the atom projected density of states in Fig.~\ref{Fig2}(b).

The topological phonon analysis of CdGeAs$_{2}$ is shown in Figs.~\ref{Fig2},~\ref{Fig3}, and~\ref{Fig4}.  The acoustic phonon modes along the in-plane $\Gamma$-$X$ direction are non-degenerate whereas, the two lowest energy acoustic modes are degenerate along the $\Gamma$ - $Z$ direction. Top transverse acoustic (TTA) mode possesses linear phonon dispersion along the in- and out-of-plane directions.

For the topological analysis, a few of the phonon bands are selected, which are highlighted in Fig.~\ref{Fig2}(a) from their bulk band crossings. Bands $8$ and $9$ show band crossings leading to the symmetry-protected nodal rings, nodal line, Weyl, and triply degenerate points inside the BZ. All the nodal phases formed by the crossings of these bands are summarized in Fig.~\ref{Fig3}(a). The Weyl points are located parallel to $\Gamma$ - $X$ directions but not at the $k_{z}$ = 0 plane. The nodal line formed from these bands is represented by the green color in Fig.~\ref{Fig2}(a), at the endpoint of the nodal ring band 7 appears to form the triply degenerate point along the $\Gamma$ - $Z$ direction. The triply degenerate points of the 7, 8, and 9 bands crossings are protected by the $C_{3z}$ rotational symmetry. The dispersion along the $k_{z}$ direction for these three bands are shown separately in Fig.~\ref{Fig3}(b).  The triply degenerate points are obtained from the crossings of doubly degenerate and non-degenerate bands. The 3D visualization of these three bands in $k_{x}-k_{y}$ and $k_{x}-k_{z}$ planes can be observed from the Fig.~\ref{Fig3}(c,d). Fig.~\ref{Fig3}(c) shows that the TDP point is formed by the crossings of two nearly flat and one dispersive band. Since the slope of the crossing bands appears to be the same, therefore, type-II TDP is resolved in the $k_x-k_y$ plane. The two TDP points along the $k_{z}$ axis are connected through a straight doubly degenerated Weyl nodal line. The calculated Berry phase for this nodal line is 0 suggesting a topologically trivial character and that is followed by the in-plane quadratic dispersion as shown in Fig.~\ref{Fig3}(e). The nodal line is formed by the bands 8 and 9 as depicted in Fig.~\ref{Fig3}(a). Below the TDP along $k_{z}$ axis two nodal rings intersect which are lying on the diagonal mirror planes and the Berry phase of both rings is 0, which again reveals the topologically trivial character of these rings and the band dispersion along the diagonal axis is depicted in Fig.~\ref{Fig3}(f). The crossing point is formed by one nearly flat and one dispersive band. For the Berry phase calculation, a circle contour is defined perpendicular to the nodal plane with a small finite radius to avoid inclosing other band crossings.  
 
Moreover, eight Weyl points are observed parallel to $k_{x}$ and $k_{y}$ axes at $k_{z}$ = $\pm$ 0.21 \AA$^{-1}$. These eight Weyl points formed by these bands possess a magnitude of charge $1$ and all the Weyl points are at the same frequency 2.41 THz as a result of mirror inversion in the $m_{xy}$ diagonal planes. For same $k_{x}$ and $k_{y}$ coordinates opposite $k_{z}$ WPs have the same chiral charge owing to the absence of the $m_{z}$ symmetry thus the projected WPs on the (001) surface possess a chiral charge of $\pm$ 2 and are at the $k_{x}$ and $k_{y}$ axes as illustrated in the Fig.~\ref{Fig3}(i). The Berry curvature around a positive chiral WP is non-zero as depicted in Fig.~\ref{Fig3}(h). Opposite chiral charge Weyl nodes connect by the topologically nontrivial arcs on (001) surface as depicted in Fig.~\ref{Fig3}(g) for phonon energy 2.44~THz.

Similar to the preceding analysis, we also analyzed topological features of the crossings of $14$ and $15$ phonon bands as identified in Fig.~\ref{Fig4}(a). The nodal line is guided in green color and a TDP is present inside the shown box. Unlike bands 8 and 9, these phonon bands have two pairs of Weyl points along the principle momentum axes at $k_{z}$ = 0 plane. Notably, the magnitude of the chiral charges remains the same as before. The nodal points of these bands are summarized in Fig.~\ref{Fig4}(a). Along $k_{z}$ axis 
a pair of TDP is observed at $k_{z}$= $\pm 0.37$  \AA$^{-1}$. These points are connected with the open straight line that possesses zero Berry phase and obeys the non-linear band characters along the in-plane direction. Whereas along the crossing line it shows the linear band crossings as represented in Fig.~\ref{Fig4}(e,f). The TDP is formed by the crossings of 14, 15, and 16 bands, whose 3D dispersion along $k_{x}k_{y}$ is shown in Fig.~\ref{Fig4}(c). Bands 15 and 16 are doubly degenerate along $k_{z}$ axis before the TDP crossings and after the crossing bands, 14 and 15 become degenerate for a range of $k$-values as shown in Fig.~\ref{Fig4}(b,d).

We also discuss the Weyl phonons which are present on the $k_{z}$ = 0 plane along the $k_{x}$ and $k_{y}$ axes at $|k_{x}| = |k_{y}| = 0.31~$ \AA$^{-1}$. The energy-momentum relations and locations of WPs in the 3D are shown in Fig.~\ref{Fig4}(g).  The Berry phases around the opposite chiral WPs are calculated using the Wilson loop method from the Wannier charges. Positive and negative chiral WPs possess $\pi$ and $-\pi$ Berry phases that can be observed from Fig.~\ref{Fig4}(h-i). WPs on $k_{x}$ axis and $k_{y}$ axis have opposite chiral charges. The opposite chiral Weyl points are connected through the arcs as depicted in Fig.~\ref{Fig4}(j) and the corresponding surface states of the opposite WPs on to the (001) surface are shown 
in Fig.~\ref{Fig4}(k,l).

The topology in the phonon bands leads to the symmetry protected states which stays robust against symmetry respected perturbations. There are many other bands in the phonon spectrum featuring topological states, for example acoustic and optical bands possess several straight nodal lines and TDP along $k_{z}$ direction protected by the crystalline symmetries. Topological protection of phonon states lead to many novel phenomena and increase in the phonon scatterings result in the reduction of the lattice thermal conductivity that leads to the thermoelectric properties. It has been established for many of the materials such as TaSb, TaBi, NbSb etc. that having the topological phonons lead to the lower lattice thermal conductivity than the materials which do not have the topological characters in the phonons~\cite{singh2018topological, li2016influence, yu2021absence}. In parallel to topological phonons, many electronic properties such as high band gap value between valence and conduction bands that suppress the bipolar effect prevents the reduction of thermopower below a certain temperature. Heavy hole band for $p$-type doping gives rise to the large thermopower, and the anisotropic nature of hole band supports to amplify the thermoelectric power factor and $ZT$ by increasing the mobility of charge carriers along highly dispersive band directions.

We present the calculated bulk band structure obtained with both the GGA and mBJ XC functionals in Figs.~\ref{Fig5}(a-c) in the presence of spin-orbit coupling (SOC). A negative bandgap of $\sim0.08$ eV is obtained with GGA XC functional with a clear band inversion at $\Gamma$ point [Figs.~\ref{Fig5}(a) and Fig.~\ref{Fig5}(b)]. The orbital resolved band structure is shown in Fig.~\ref{Fig5}(b) where As $p$ (blue color) and Ge $s$ (green color) states contribute dominantly near the Fermi level. A band inversion between the As $p$ and Ge $s$ states is evident at the $\Gamma$-point. We also examine the electronic structure in the absence of SOC.   

\begin{table}
\caption{\label{Tab1}Location of Weyl nodes on the $k_{z}= 0 $ plane without SOC obtained with GGA XC functional.} 
\begin{ruledtabular}
\begin{tabular}{ccccc}
			%\begin{tabular}{c |c |c |c |c |c |c |c |c |c |c|c} 
			
			%	\vspace{0.25 cm}
			Strain & $\rm b^{'}$$=$b	& 1.01b &  1.02b & 1.03b \\  \\
			Weyl position & &  &  & \\  
			$k_{x}$ $=$ $k_{y}$ ($\rm \AA^{-1}$) & 0.028   & 0.022     &  0.016  & 0.008  \\
			\\
			%\vspace{0.25 cm}
			%\hline
			
		\end{tabular}
	\end{ruledtabular}
\end{table}  

Weyl points of positive chiral charges are located at $k_{x}$ axis and negative charges are found at $k_{y}$ axis. For surface states, four Weyl points in the bulk are projected onto (001) surface, and two opposite chiral Weyl points are connected with the arcs as represented in Fig.~\ref{Fig5}(g). To check the effect of strain on the location of Weyl points in momentum space we summarized a few of the strains in Table~\ref{Tab1}. Keeping the unit cell volume constant, compression on the small $a$-axis of the tetragonal structure is applied in the form of the tensile 
strain on the longest $b$-axis by 1 to 3$\%$. In absence of SOC, the Weyl phase is robust with four Weyl nodes on the $k_{z}$ $=$ 0 plane. The separation between the Weyl nodes points continuously decreases as the strain is increased from 1$\%$ to 3$\%$ as tabulated in Table~\ref{Tab1}.

GGA XC functional underestimates band gap in the materials therefore we have carried out thermoelectric calculations in presence of the mBJ XC functional. 

Figure~\ref{Fig5}(c) shows the band structure obtained with mBJ XC functional. The band inversion is now disappeared and the system becomes trivial with a direct bandgap of 0.65 eV. This is close to the experimentally reported value of 0.57 eV at room temperature~\cite{mccrae1997photoluminescence,akimchenko1973electroreflection, bai2005urbach}. Fig.~\ref{Fig5}(d) represents the density of states (DOS) for GGA and mBJ XC functionals in red and blue colors, respectively. The DOS below the Fermi energy shows a rapid increase with energy which leads to the large thermopower for $p$-type doping as discussed below.
	
\begin{figure*}%[!]
\includegraphics[width=1.0\textwidth]{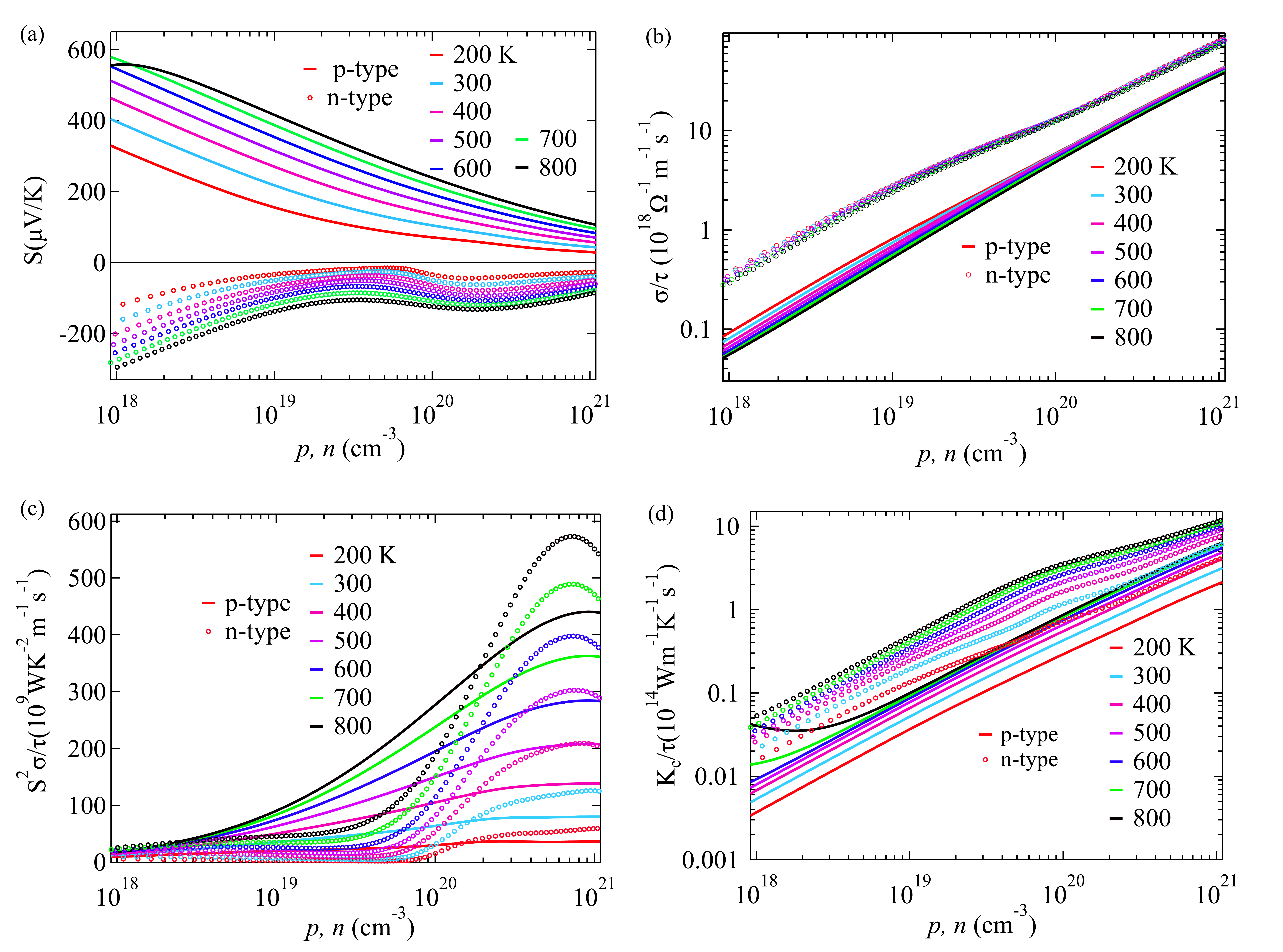}
\caption{Calculated thermoelectric properties of CdGeAs$_{2}$  for $p$-type and $n$-type doping. (a) Thermopower plot with varying carrier density at constant temperatures. (b) Electrical conductivity divided by relaxation time as a function of carrier density in the regime of $10^{18}$-$10^{21}$ cm$^{-3}$. (c) The power factor divided by relaxation time with carrier density at constant temperatures. (d)  Electrical thermal conductivity over relaxation time plotted against carrier density for various temperatures.} 
\label{Fig6}
\end{figure*}  
	
\begin{figure}%[!]
\includegraphics[width=0.48\textwidth]{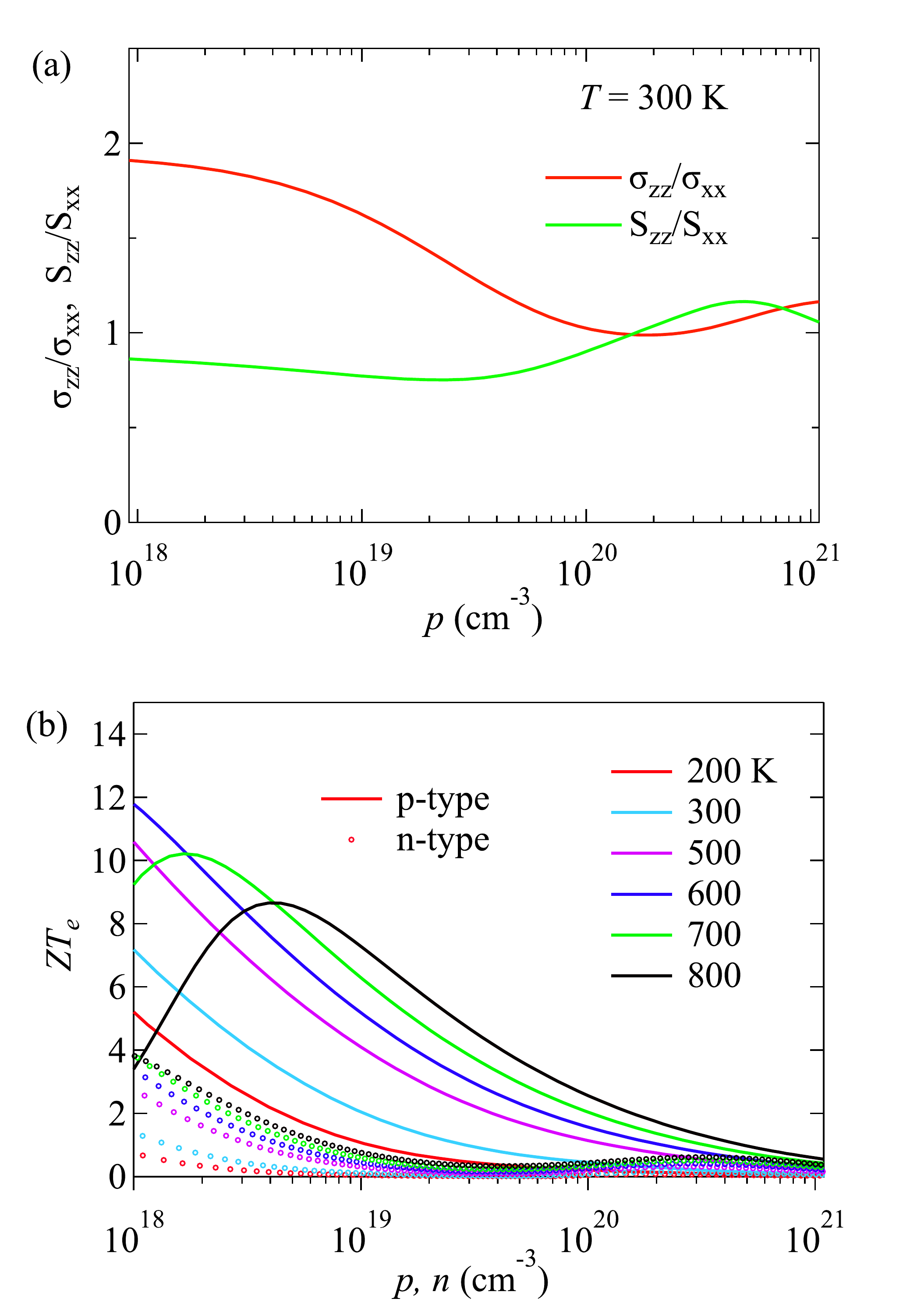}
\caption{(a) Anisotropic parameters $\sigma_{zz}/\sigma_{xx}$ and $S_{zz}/S_{xx}$ plotted as a function of hole carrier density at $T$= 300~K. (b) The upper limit of the figure-of-merit ($ZT_{e}$) as a function of carrier density at constant temperatures.}
\label{Fig7}
\end{figure}  
	
\begin{figure*}%[!]
\includegraphics[width=0.9\textwidth]{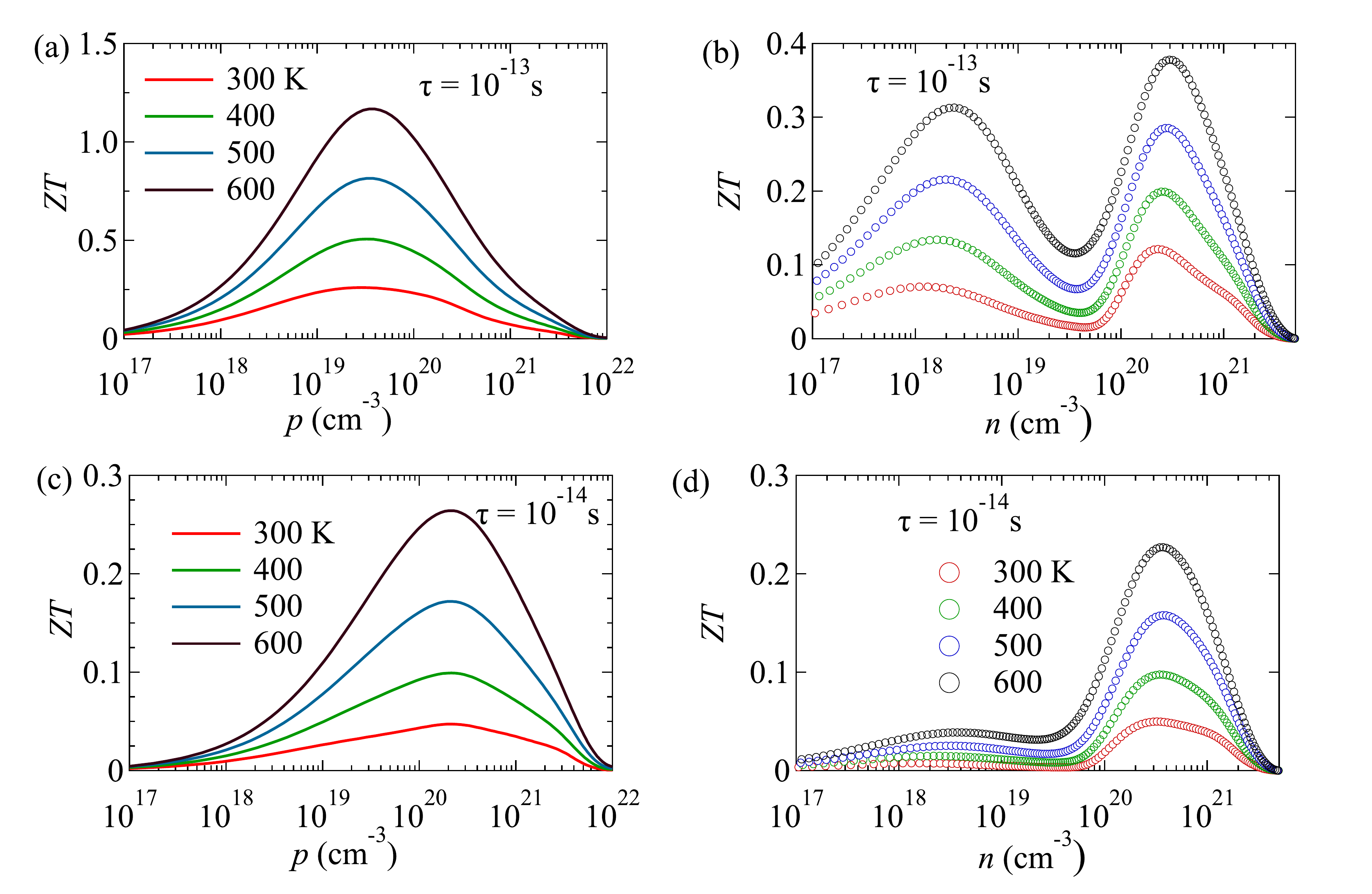}
\caption{Calculated figure-of-merit $ZT$ for CdGeAs$_{2}$. (a-b) $ZT$ values as the functions of hole and electron carrier densities at a constant scattering time $\tau$ $=$ 10$^{-13}$. (c-d) Estimation of $ZT$ values at $\tau$ $=$ 10$^{-14}$ s, for hole and electron type of carriers, respectively.}\label{Fig8}	
\end{figure*}
	
Our first-principles results show heavy hole bands along $\Gamma$-$X$ direction whereas the bands are highly dispersive along $\Gamma$-$Z$ direction [Fig.~\ref{Fig5}(e)]. This suggests that the charge carriers of in-plane direction ($\Gamma$-$X$) possesses high effective mass compared to the out-of-plane direction ($\Gamma$-$Z$) which possesses light effective mass and high mobility. To see the anisotropic effects, the weighted mobility can be defined as $\mu_{w}$ = $\mu$($\frac{m^{*}}{m_{e}})^{\frac{3}{2}}$, where 
$\mu$ is the mobility of charge carriers, $m^{*}$ = $N_{v}^{\frac{2}{3}}$ ${(m_{x}m_{y}m_{z})}^\frac{1}{3}$;  $N_{v}$ valley degeneracy, and $m_{e}$ is the rest mass of the electron.
The highly dispersive band along out-of-plane ($\Gamma$-$Z$) direction gives rise to the higher weighted mobility $\mu_{w}$ because of high mobility as compared to the in-plane $\Gamma$-$X$ direction. The high value of the $\mu_{w}$ lead to the higher power factor 
and $ZT$ parameters. The anisotropy in effective masses is calculated from transport parameters 
at room temperature $\frac{1}{m^{*}_{\alpha\beta}}$ $=$ $\frac{\sigma_{\alpha\beta}/ \tau}{e^{2}n}$; where $\sigma_{\alpha\beta} / \tau $ 
is electrical conductivity tensor divided by relaxation time, $n$ is the carrier density. Further, the anisotropic band structure is 
consistent with the calculated effective masses for $ab$-plane ($m_{xx}$) and $c$-direction ($m_{zz}$) for $p$-type CdGeAs$_{2}$ as shown 
in Fig.~\ref{Fig5}(f). In the low carrier density $p$ $\sim$ 1.27~$\times$ $10^{18}$ cm$^{-3}$ the anisotropy in the masses is maximum 
$m_{xx}$ $\sim$ 1.9~$m_{zz}$, and as the hole carrier density increases the anisotropy decreases and the effective masses for in- and 
out-of-plane bands are around $m_{xx}$ = 0.73~$m_{e}$ and $m_{zz}$ = 0.63~$m_{e}$ for $p$ $\sim$ $10^{21}$~cm$^{-3}$.       
	
Figures~\ref{Fig6},~\ref{Fig7}, and ~\ref{Fig8} show the calculated thermoelectric properties of CdGeAs$_2$ within constant relaxation 
time approximation~\cite{singh2010doping,parker2010high,wang2011heavily,sun2016thermoelectric,sun2016ther}. The thermopower ($S$) of 
$p$-type carriers (solid curves) is greater than $n$-type carriers (dotted curves) in the entire carrier density range 
10$^{18}$ - 10$^{21}$ cm$^{-3}$ at a constant temperature as shown in  Fig.~\ref{Fig6}(a). The thermopower increases with increasing temperature and reaches a value of 567 $\mu$V/K at 700 K for $p$-type doping and 270 $\mu$V/K 
at 800 K for $n$-type at carrier concentration 10$^{18}$ cm$^{-3}$.
	
To understand the large thermopower of $p$-type carriers, we have used the Mott relation in constant relaxation time approximation 
which is given by~\cite{he2017advances, irkhin2007electronic}.
	\begin{equation}
		S(n,T) =  \frac{\pi^{2} k_{B}^{2}T}{3q} \left[\frac{1}{n}\frac{dn(E)}{dE} + \frac{1}{\mu}\frac{d\mu(E)}{dE}\right]_{E_{F}} 
		\label{Eq1}
	\end{equation} 
Here, $q$ denotes the charge, $n(E)$ represents the density of states, $\mu$ is the mobility, and $T$ denotes the absolute temperature. The equation is composed of the addition of two derivative terms. It can be simply related from the first term of Eqn.~\ref{Eq1} that steepness in the DOS enhances thermopower. The mBJ XC calculated DOS (Fig~\ref{Fig5}(d)) for $p$-type doping shows a steep rise than the $n$-type doping in CdGeAs$_{2}$ resulting large thermopower for the $p$ doped system.  The second term of the Mott relation exhibits that if the mobility of charge carriers increases with energy as a consequence of the critical scatterings near the Fermi level then the thermopower can be boosted further.
	
Thermopower of both types of carriers gradually reduce with increasing carrier density except for the $p$-type 800 K plot. For $p$-type carriers, owing to electronic thermal excitations, the thermopower reduces with decreasing carrier density at high temperatures as observed in Fig.~\ref{Fig6}(a) for 800 K.
	
The bipolar conduction effect in the low carrier density regime for the narrow gap semiconductors is kind of normal, as studies suggest, to suppress this effect the band gap energy should be more than around 8~$k_{\rm B}$$T$~\cite{dehkordi2015thermoelectric}.
	
The band gap of CdGeAs$_{2}$ (0.65 eV) is reasonably large which ensures that the bipolar effect is not seen at temperatures below 700 K. The band gap of CdGeAs$_{2}$ is higher than the calculated values for the other TE materials such as (PbSe ($E_{g}$ = 0.28) and PbTe ($E_{g}$ = 0.36) from ref.~\cite{ekuma2012optical}). Also, thermopower at 300 K for the $p$-type CdGeAs$_{2}$ is more than the PbSe, SnTe and PbTe compounds~\cite{singh2010doping,singh2010thermopower, parker2010high}. 
	
The electrical conductivity divided by the relaxation time $\frac{\sigma}{\tau}$ is represented in Fig.~\ref{Fig6}(b) 
on a logarithmic scale for both types of charge carriers for different temperatures from 200 to 800~K. The $\frac{\sigma}{\tau}$ of the
$n$-type charge carriers is higher than the $p$-type carriers that 
is consistent from the highly dispersive nature of the $n$-type band compared to the $p$-type band that results to have relatively small effective mass of $n$-type band which essentially increases the electrical conductivity. The increase in temperature reduces the electrical conductivity in the entire range of carrier density  10$^{18}$ - 10$^{21}$ cm$^{-3}$. However, the magnitude of change with the temperature at a given carrier concentration is not much for both types of carriers. For $p$-type charge carriers $\frac{\sigma}{\tau}$ is (0.1 and 0.06) $\times$ 10$^{18}$ $\Omega^{-1}$ $m^{-1}$ s$^{-1}$ at carrier concentration 10$^{18}$ cm$^{-3}$ and for temperatures 300 and 800~K. 
	
The power factor divided by the relaxation time $\frac{S^{2}\sigma}{\tau}$ is shown in Fig.~\ref{Fig6}(c) for the both types of charge 
carriers in the range of carrier density 10$^{18}$ - 10$^{21}$ cm$^{-3}$. It is obvious from the figure that the
$p$-type power factor is dominated over the $n$-type power factor in intermediate carrier density regime before crossover appears around 
1.07$\times$10$^{20}$ cm$^{-3}$ and after the crossover $n$-type power factor is dominated over $p$-type until carrier density reaches 
up to 10$^{21}$ cm$^{-3}$ for each temperature ranging from 200 to 800~K. 
	
The power factor is one of the considerable parameters to increase the thermoelectric performance of materials. However, for the rough estimation of power factor if empirically relaxation time is considered in the order of 10$^{-14}$~s then it governs reasonably good values of the power factor in the range of mWK$^{-2}$m$^{-1}$ to $\mu$WK$^{-2}$m$^{-1}$ for the $p$-type carriers from high to low density regime, attributed to the optimized band structure of CdGeAs$_{2}$. 
	
The anisotropic ratio of out-of- and in-plane thermopower $S_{zz}/S_{xx}$ and electrical conductivity $\sigma_{zz}/\sigma_{xx}$ are plotted 
against hole carrier density at $T$= 300 K in the Fig.~\ref{Fig7}(a). The thermopower along the out-of-plane direction is lower than the 
in-plane direction as a consequence of light band along $\Gamma$-$Z$ direction which leads to the small effective mass. The  
$S_{zz}/S_{xx}$ is weakly changing in the low carrier density regime.
Conversely, the electrical conductivity for out-of-plane direction is $\sigma_{zz}$~$\sim$~1.9~$\sigma_{xx}$ at carrier density 
$p$~$\sim$~1.27$\times$10$^{18}$~cm$^{-3}$. Beyond this hole density, the anisotropy in conductivity decreases up to 
1.9$\times$10$^{20}$~cm$^{-3}$ and later this increases very gradually which attains $\sigma_{zz}$~$\sim$~1.16~$\sigma_{xx}$ at carrier 
density $p$~$\sim$~10$^{21}$~cm$^{-3}$. The anisotropy in the calculated transport parameters is consistent with the anisotropic 
effective mass as discussed in Fig.~\ref{Fig5}(f). 
	
Finally, the anisotropic nature of the hole band at $\Gamma$ point is shown in Fig.~\ref{Fig5} (h) where the Fermi level is shifted below 
$0.2$ eV from its pristine value which drives the anisotropy in the electrical conductivity that helps to increase the weighted mobility $\mu_{w}$ for  out-of-plane ($\Gamma$-$Z$) direction thus resulting in the high power factor and $ZT$ for out-of-plane direction compared to in-plane direction. For comparison, we also have the lower conduction band visualization while the Fermi level is around 1~eV up from the maxima of the valence band that shows the isotropic nature of electron band. Therefore, anisotropic advantage 
of the hole bands may enhance the thermoelectric properties of $p$-type carriers compared to the $n$-type carriers.         

The electronic thermal conductivity ($k_{e}$) can be defined as $k_e$$=$ L$\sigma$T. The calculated $k_{e}$ is shown in Fig.~\ref{Fig6}(d) for both carriers in the temperature range from 200 to 800~K. The electronic thermal conductivity divided by relaxation time $\frac{k_{e}}{\tau}$ for $p$-type doping is smaller than the $n$-type doping in the entire carrier density regime $10^{18} - 10^{21}$~cm$^{-3}$ at the constant temperatures which supports the higher $ZT{_e}$ values compared to $n$-type doping as shown in Fig.~\ref{Fig7}(b). 
	
For $n$-type doping, $\frac{k_{e}}{\tau}$ increases gradually with increasing temperature in the entire carrier density range  $10^{18} - 10^{21}$~cm$^{-3}$ and for $p$-type doping $\frac{k_{e}}{\tau}$ increases gradually with increasing temperature in the entire carrier density regime below 700~K. At 700~K and above in low density regime 10$^{18}$ - 10$^{19}$~cm$^{-3}$ electronic thermal conductivity increases much more rapidly with lowering the carrier density as shown in Fig.~\ref{Fig6}(d) which results in the reduction of $ZT_{e}$ values for $T\gtrsim700$~K (Fig.~\ref{Fig7}(b)).
	
The upper limit of figure-of-merit $ZT{_e}$ $=$ $\frac{S^{2}\sigma/\tau}{k_{e}/\tau}$ in the low-carrier carrier density regime are 
exceptionally high for $p$-type carriers (11.5 at $T= 600$~K and $n$ $\sim$ 1.1$\times$ 10$^{18}$~cm$^{-3}$) than the $n$-type 
carriers (3.1 at $T=600$~K and $n$ $\sim$ 1.1$\times$ 10$^{18}$~cm$^{-3}$) as a consequence of optimized band structure as depicted 
in Fig.~\ref{Fig7}(b).	
	
As the phonon dispersion of CdGeAs$_{2}$ shown in Fig.~\ref{Fig2}(a) uncover that the small frequencies of acoustic phonons attribute to 
the low group velocity which can lead to low lattice thermal conductivity ($k_l$ $=$ $ \frac{1}{3}C_{V}v_{g}l$). Moreover, as we discussed 
above the mixing of acoustic and optical modes along with the topological protection give rise the enhanced phonon scatterings resulting 
into small mean free path and thus indicate a small lattice thermal conductivity in CdGeAs$_{2}$.
	
However, the experimentally observed lattice thermal conductivity of CdGeAs$_{2}$ at 300 K is 4~Wm$^{-1}$K$^{-1}$ ~\cite{spitzer1970lattice}.
As we discussed the electronic part of the figure-of-merit $ZT_{e}$ is remarkably high for $p$-type CdGeAs$_{2}$ and notably this does not 
depend on the relaxation time $\tau$. The thermoelectric figure-of-merit can be written in terms of electronic part of figure-of-merit 
as $ZT$ $=$ $\frac{S^2\sigma}{k_{e}+k_{l}}$= $\frac{ZT_{e}}{(1+k_{l}/k_{e})}$. To estimate $ZT$ we need to have the electronic thermal 
conductivity $k_{e}$. Since from the calculations, we obtain $\frac{k_{e}}{\tau}$ therefore to get $k_{e}$ the knowledge of relaxation 
time $\tau$ is required. For the materials, $\tau$ can be a function of temperature and carrier density, at the simplest electron-phonon 
scatterings show the inverse temperature behavior $\tau$ $\propto$ $T$$^{-1}$. To get the idea of $ZT$ values in $p$- and $n$-type 
CdGeAs$_{2}$, we approximate the relaxation time as 10$^{-13}$ and 10$^{-14}$ s independent of temperature and carrier density and 
later we will add the effect of these parameters on $\tau$ for $ZT$ values.
	
Figure~\ref{Fig8} depicts the calculated $ZT$ for $\tau$ = 10$^{-13}$ and 10$^{-14}$ s. For $\tau$ $=$ 10$^{-13}$ s $ZT$ reaches from 
0.25 at 300~K to 1.16 at 600~K for $p$-type dopings as shown in Fig.~\ref{Fig8}(a). Whereas it goes up to 0.37 at 600 K from 0.12 at 300 K, 
there is one thing to note that we have used the temperature independent lattice thermal conductivity $k_{l}$ as 4~W/m-K but 
conventionally $k_{l}$ decreases with increasing temperature, therefore, we may expect further increase in the $ZT$ values.
	
For the $\tau$ $=$ 10$^{-14}$ s maximum values of $ZT$ are around 0.047 and 0.049 obtained at 300~K, respectively for $p$- and $n$-type 
dopings, and with increasing temperature $ZT$ increases and attains 0.26 and 0.22 at the 600~K for $p$- and $n$-type dopings, respectively. 
Similar to the previous discussion, drops in the $k_{l}$ with increasing temperature may be expected to enhance the $ZT$ values further. Of 
course with increasing temperature $\tau$ reduces and without experimental data the scaling of carrier density is challenging but as the 
previous studies suggest this makes $\tau$ to be suppressed  ~\cite{sun2016thermoelectric}. Thus incorporating both of the these effects 
into the $\tau$ would lead to the decrease in the estimated values of $ZT$.
	
These analytical results indicate towards a reasonably good $ZT$ values at high temperatures $T$ $\gtrsim$ 300~K. Experimental engineering of the reinforcing phonon scatterings from defects and grain boundaries will further reduce the $k_{l}$ and enhancement of the $k_{e}$ would increase the $ZT$ values for the better performance of thermoelectric devices. In this work, we have covered the complete comprehensive study of the CdGeAs$_{2}$ from the calculated results these will be surely helpful for the further leads to explore the experimental aspects. 
	
\section*{Summary and Conclusion}
We have studied the topological states in the phonon spectrum and resolved the triply degenerate phonons on the $k_{z}$ axis. Multiple pairs of Weyl points in the bulk structure, and their topological surface states are observed on the (001) surface. The topological features in the phonon spectra could suppress lattice thermal conductivity enhancing the thermoelectric figure of merit. We discuss electronic properties of CdGeAs$_{2}$ that exhibit many thermoelectric supportive features and result in high thermopower for $p$-type dopings. The anisotropic hole bands lead to the high-weighted mobility that supports the high power factor and $ZT$ values. The calculated value of electronic figure-of-merit $ZT_{e}$ is high (more than 2 for temperatures 300~K and above) in carrier density regime 10$^{18}$ - 10$^{19}$ cm$^{-3}$ for $p$-type CdGeAs${_2}$ that result in the noble response to the thermoelectric performance. Our study unfold that CdGeAs${_2}$ has topologically nontrivial phonon states and exhibits very good theromelectric properties that are surely useful for the potential applications.   
	
\section*{Acknowledgement}
We thank Prof. Kalobaran Maiti for providing the computational resources and acknowledge the TIFR computing resources. This work was supported by the Department of Atomic Energy of the government of India under Project No. 12-R$\&$D-TFR-5.10-0100. 	

\section*{Data Availability}
The data supporting the findings of this study are available within the article. More data can be provided on a reasonable request to the corresponding author. 
	
	%% If you have bibdatabase file and want bibtex to generate the
	%% bibitems, please use
	%%
%\bibliographystyle{elsarticle-num} 
%\bibliography{references}

\begin{thebibliography}{69}%
	\makeatletter
	\providecommand \@ifxundefined [1]{%
		\@ifx{#1\undefined}
	}%
	\providecommand \@ifnum [1]{%
		\ifnum #1\expandafter \@firstoftwo
		\else \expandafter \@secondoftwo
		\fi
	}%
	\providecommand \@ifx [1]{%
		\ifx #1\expandafter \@firstoftwo
		\else \expandafter \@secondoftwo
		\fi
	}%
	\providecommand \natexlab [1]{#1}%
	\providecommand \enquote  [1]{``#1''}%
	\providecommand \bibnamefont  [1]{#1}%
	\providecommand \bibfnamefont [1]{#1}%
	\providecommand \citenamefont [1]{#1}%
	\providecommand \href@noop [0]{\@secondoftwo}%
	\providecommand \href [0]{\begingroup \@sanitize@url \@href}%
	\providecommand \@href[1]{\@@startlink{#1}\@@href}%
	\providecommand \@@href[1]{\endgroup#1\@@endlink}%
	\providecommand \@sanitize@url [0]{\catcode `\\12\catcode `\$12\catcode
		`\&12\catcode `\#12\catcode `\^12\catcode `\_12\catcode `\%12\relax}%
	\providecommand \@@startlink[1]{}%
	\providecommand \@@endlink[0]{}%
	\providecommand \url  [0]{\begingroup\@sanitize@url \@url }%
	\providecommand \@url [1]{\endgroup\@href {#1}{\urlprefix }}%
	\providecommand \urlprefix  [0]{URL }%
	\providecommand \Eprint [0]{\href }%
	\providecommand \doibase [0]{https://doi.org/}%
	\providecommand \selectlanguage [0]{\@gobble}%
	\providecommand \bibinfo  [0]{\@secondoftwo}%
	\providecommand \bibfield  [0]{\@secondoftwo}%
	\providecommand \translation [1]{[#1]}%
	\providecommand \BibitemOpen [0]{}%
	\providecommand \bibitemStop [0]{}%
	\providecommand \bibitemNoStop [0]{.\EOS\space}%
	\providecommand \EOS [0]{\spacefactor3000\relax}%
	\providecommand \BibitemShut  [1]{\csname bibitem#1\endcsname}%
	\let\auto@bib@innerbib\@empty
	%</preamble>
	\bibitem [{\citenamefont {Singh}\ \emph {et~al.}()\citenamefont {Singh},
		\citenamefont {Lin},\ and\ \citenamefont {Bansil}}]{AMPER2022_BS}%
	\BibitemOpen
	\bibfield  {author} {\bibinfo {author} {\bibfnamefont {B.}~\bibnamefont
			{Singh}}, \bibinfo {author} {\bibfnamefont {H.}~\bibnamefont {Lin}},\ and\
		\bibinfo {author} {\bibfnamefont {A.}~\bibnamefont {Bansil}},\ }\href
	{https://doi.org/https://doi.org/10.1002/adma.202201058} {\bibinfo  {journal}
		{Adv. Mater.}\ ,\ \bibinfo {pages} {2201058}}\BibitemShut {NoStop}%
	\bibitem [{\citenamefont {Bansil}\ \emph {et~al.}(2016)\citenamefont {Bansil},
		\citenamefont {Lin},\ and\ \citenamefont {Das}}]{RMP_Bansil}%
	\BibitemOpen
	\bibfield  {journal} {  }\bibfield  {author} {\bibinfo {author} {\bibfnamefont
			{A.}~\bibnamefont {Bansil}}, \bibinfo {author} {\bibfnamefont
			{H.}~\bibnamefont {Lin}},\ and\ \bibinfo {author} {\bibfnamefont
			{T.}~\bibnamefont {Das}},\ }\href
	{https://doi.org/10.1103/RevModPhys.88.021004} {\bibfield  {journal}
		{\bibinfo  {journal} {Rev. Mod. Phys.}\ }\textbf {\bibinfo {volume} {88}},\
		\bibinfo {pages} {021004} (\bibinfo {year} {2016})}\BibitemShut {NoStop}%
	\bibitem [{\citenamefont {Hasan}\ and\ \citenamefont {Kane}(2010)}]{RMP_Hasan}%
	\BibitemOpen
	\bibfield  {author} {\bibinfo {author} {\bibfnamefont {M.~Z.}\ \bibnamefont
			{Hasan}}\ and\ \bibinfo {author} {\bibfnamefont {C.~L.}\ \bibnamefont
			{Kane}},\ }\href {https://doi.org/10.1103/RevModPhys.82.3045} {\bibfield
		{journal} {\bibinfo  {journal} {Rev. Mod. Phys.}\ }\textbf {\bibinfo {volume}
			{82}},\ \bibinfo {pages} {3045} (\bibinfo {year} {2010})}\BibitemShut
	{NoStop}%
	\bibitem [{\citenamefont {Liu}\ \emph {et~al.}(2014)\citenamefont {Liu},
		\citenamefont {Zhou}, \citenamefont {Zhang}, \citenamefont {Wang},
		\citenamefont {Weng}, \citenamefont {Prabhakaran}, \citenamefont {Mo},
		\citenamefont {Shen}, \citenamefont {Fang}, \citenamefont {Dai} \emph
		{et~al.}}]{liu2014discovery}%
	\BibitemOpen
	\bibfield  {author} {\bibinfo {author} {\bibfnamefont {Z.}~\bibnamefont
			{Liu}}, \bibinfo {author} {\bibfnamefont {B.}~\bibnamefont {Zhou}}, \bibinfo
		{author} {\bibfnamefont {Y.}~\bibnamefont {Zhang}}, \bibinfo {author}
		{\bibfnamefont {Z.}~\bibnamefont {Wang}}, \bibinfo {author} {\bibfnamefont
			{H.}~\bibnamefont {Weng}}, \bibinfo {author} {\bibfnamefont {D.}~\bibnamefont
			{Prabhakaran}}, \bibinfo {author} {\bibfnamefont {S.-K.}\ \bibnamefont {Mo}},
		\bibinfo {author} {\bibfnamefont {Z.}~\bibnamefont {Shen}}, \bibinfo {author}
		{\bibfnamefont {Z.}~\bibnamefont {Fang}}, \bibinfo {author} {\bibfnamefont
			{X.}~\bibnamefont {Dai}}, \emph {et~al.},\ }\href@noop {} {\bibfield
		{journal} {\bibinfo  {journal} {Science}\ }\textbf {\bibinfo {volume}
			{343}},\ \bibinfo {pages} {864} (\bibinfo {year} {2014})}\BibitemShut
	{NoStop}%
	\bibitem [{\citenamefont {Jenkins}\ \emph {et~al.}(2016)\citenamefont
		{Jenkins}, \citenamefont {Lane}, \citenamefont {Barbiellini}, \citenamefont
		{Sushkov}, \citenamefont {Carey}, \citenamefont {Liu}, \citenamefont
		{Krizan}, \citenamefont {Kushwaha}, \citenamefont {Gibson}, \citenamefont
		{Chang} \emph {et~al.}}]{jenkins2016three}%
	\BibitemOpen
	\bibfield  {author} {\bibinfo {author} {\bibfnamefont {G.}~\bibnamefont
			{Jenkins}}, \bibinfo {author} {\bibfnamefont {C.}~\bibnamefont {Lane}},
		\bibinfo {author} {\bibfnamefont {B.}~\bibnamefont {Barbiellini}}, \bibinfo
		{author} {\bibfnamefont {A.}~\bibnamefont {Sushkov}}, \bibinfo {author}
		{\bibfnamefont {R.}~\bibnamefont {Carey}}, \bibinfo {author} {\bibfnamefont
			{F.}~\bibnamefont {Liu}}, \bibinfo {author} {\bibfnamefont {J.}~\bibnamefont
			{Krizan}}, \bibinfo {author} {\bibfnamefont {S.}~\bibnamefont {Kushwaha}},
		\bibinfo {author} {\bibfnamefont {Q.}~\bibnamefont {Gibson}}, \bibinfo
		{author} {\bibfnamefont {T.-R.}\ \bibnamefont {Chang}}, \emph {et~al.},\
	}\href@noop {} {\bibfield  {journal} {\bibinfo  {journal} {Phys. Rev. B}\
		}\textbf {\bibinfo {volume} {94}},\ \bibinfo {pages} {085121} (\bibinfo
		{year} {2016})}\BibitemShut {NoStop}%
	\bibitem [{\citenamefont {Nie}\ \emph {et~al.}(2018)\citenamefont {Nie},
		\citenamefont {Meng}, \citenamefont {Li}, \citenamefont {Luan},\ and\
		\citenamefont {Yu}}]{nie2018phase}%
	\BibitemOpen
	\bibfield  {author} {\bibinfo {author} {\bibfnamefont {T.}~\bibnamefont
			{Nie}}, \bibinfo {author} {\bibfnamefont {L.}~\bibnamefont {Meng}}, \bibinfo
		{author} {\bibfnamefont {Y.}~\bibnamefont {Li}}, \bibinfo {author}
		{\bibfnamefont {Y.}~\bibnamefont {Luan}},\ and\ \bibinfo {author}
		{\bibfnamefont {J.}~\bibnamefont {Yu}},\ }\href@noop {} {\bibfield  {journal}
		{\bibinfo  {journal} {J. Condens. Matter Phys.}\ }\textbf {\bibinfo {volume}
			{30}},\ \bibinfo {pages} {125502} (\bibinfo {year} {2018})}\BibitemShut
	{NoStop}%
	\bibitem [{\citenamefont {Wang}\ \emph {et~al.}(2012)\citenamefont {Wang},
		\citenamefont {Sun}, \citenamefont {Chen}, \citenamefont {Franchini},
		\citenamefont {Xu}, \citenamefont {Weng}, \citenamefont {Dai},\ and\
		\citenamefont {Fang}}]{wang2012dirac}%
	\BibitemOpen
	\bibfield  {author} {\bibinfo {author} {\bibfnamefont {Z.}~\bibnamefont
			{Wang}}, \bibinfo {author} {\bibfnamefont {Y.}~\bibnamefont {Sun}}, \bibinfo
		{author} {\bibfnamefont {X.-Q.}\ \bibnamefont {Chen}}, \bibinfo {author}
		{\bibfnamefont {C.}~\bibnamefont {Franchini}}, \bibinfo {author}
		{\bibfnamefont {G.}~\bibnamefont {Xu}}, \bibinfo {author} {\bibfnamefont
			{H.}~\bibnamefont {Weng}}, \bibinfo {author} {\bibfnamefont {X.}~\bibnamefont
			{Dai}},\ and\ \bibinfo {author} {\bibfnamefont {Z.}~\bibnamefont {Fang}},\
	}\href@noop {} {\bibfield  {journal} {\bibinfo  {journal} {Phys. Rev. B}\
		}\textbf {\bibinfo {volume} {85}},\ \bibinfo {pages} {195320} (\bibinfo
		{year} {2012})}\BibitemShut {NoStop}%
	\bibitem [{\citenamefont {Sun}\ \emph {et~al.}(2015{\natexlab{a}})\citenamefont
		{Sun}, \citenamefont {Wu},\ and\ \citenamefont {Yan}}]{sun2015topological}%
	\BibitemOpen
	\bibfield  {author} {\bibinfo {author} {\bibfnamefont {Y.}~\bibnamefont
			{Sun}}, \bibinfo {author} {\bibfnamefont {S.-C.}\ \bibnamefont {Wu}},\ and\
		\bibinfo {author} {\bibfnamefont {B.}~\bibnamefont {Yan}},\ }\href@noop {}
	{\bibfield  {journal} {\bibinfo  {journal} {Phys. Rev. B}\ }\textbf {\bibinfo
			{volume} {92}},\ \bibinfo {pages} {115428} (\bibinfo {year}
		{2015}{\natexlab{a}})}\BibitemShut {NoStop}%
	\bibitem [{\citenamefont {Yan}\ and\ \citenamefont
		{Felser}(2017)}]{yan2017topological}%
	\BibitemOpen
	\bibfield  {author} {\bibinfo {author} {\bibfnamefont {B.}~\bibnamefont
			{Yan}}\ and\ \bibinfo {author} {\bibfnamefont {C.}~\bibnamefont {Felser}},\
	}\href@noop {} {\bibfield  {journal} {\bibinfo  {journal} {Annu. Rev.
				Condens.}\ }\textbf {\bibinfo {volume} {8}},\ \bibinfo {pages} {337}
		(\bibinfo {year} {2017})}\BibitemShut {NoStop}%
	\bibitem [{\citenamefont {Xu}\ \emph {et~al.}(2015)\citenamefont {Xu},
		\citenamefont {Alidoust}, \citenamefont {Belopolski}, \citenamefont {Zhang},
		\citenamefont {Bian}, \citenamefont {Chang}, \citenamefont {Zheng},
		\citenamefont {Strokov}, \citenamefont {Sanchez}, \citenamefont {Chang} \emph
		{et~al.}}]{xu2015discovery}%
	\BibitemOpen
	\bibfield  {author} {\bibinfo {author} {\bibfnamefont {S.-Y.}\ \bibnamefont
			{Xu}}, \bibinfo {author} {\bibfnamefont {N.}~\bibnamefont {Alidoust}},
		\bibinfo {author} {\bibfnamefont {I.}~\bibnamefont {Belopolski}}, \bibinfo
		{author} {\bibfnamefont {C.}~\bibnamefont {Zhang}}, \bibinfo {author}
		{\bibfnamefont {G.}~\bibnamefont {Bian}}, \bibinfo {author} {\bibfnamefont
			{T.-R.}\ \bibnamefont {Chang}}, \bibinfo {author} {\bibfnamefont
			{H.}~\bibnamefont {Zheng}}, \bibinfo {author} {\bibfnamefont
			{V.}~\bibnamefont {Strokov}}, \bibinfo {author} {\bibfnamefont {D.~S.}\
			\bibnamefont {Sanchez}}, \bibinfo {author} {\bibfnamefont {G.}~\bibnamefont
			{Chang}}, \emph {et~al.},\ }\href@noop {} {\bibfield  {journal} {\bibinfo
			{journal} {arXiv preprint arXiv:1504.01350}\ } (\bibinfo {year}
		{2015})}\BibitemShut {NoStop}%
	\bibitem [{\citenamefont {Lv}\ \emph {et~al.}(2015)\citenamefont {Lv},
		\citenamefont {Muff}, \citenamefont {Qian}, \citenamefont {Song},
		\citenamefont {Nie}, \citenamefont {Xu}, \citenamefont {Richard},
		\citenamefont {Matt}, \citenamefont {Plumb}, \citenamefont {Zhao} \emph
		{et~al.}}]{lv2015observation}%
	\BibitemOpen
	\bibfield  {author} {\bibinfo {author} {\bibfnamefont {B.}~\bibnamefont
			{Lv}}, \bibinfo {author} {\bibfnamefont {S.}~\bibnamefont {Muff}}, \bibinfo
		{author} {\bibfnamefont {T.}~\bibnamefont {Qian}}, \bibinfo {author}
		{\bibfnamefont {Z.}~\bibnamefont {Song}}, \bibinfo {author} {\bibfnamefont
			{S.}~\bibnamefont {Nie}}, \bibinfo {author} {\bibfnamefont {N.}~\bibnamefont
			{Xu}}, \bibinfo {author} {\bibfnamefont {P.}~\bibnamefont {Richard}},
		\bibinfo {author} {\bibfnamefont {C.~E.}\ \bibnamefont {Matt}}, \bibinfo
		{author} {\bibfnamefont {N.~C.}\ \bibnamefont {Plumb}}, \bibinfo {author}
		{\bibfnamefont {L.}~\bibnamefont {Zhao}}, \emph {et~al.},\ }\href@noop {}
	{\bibfield  {journal} {\bibinfo  {journal} {Phys. Rev. Lett.}\ }\textbf
		{\bibinfo {volume} {115}},\ \bibinfo {pages} {217601} (\bibinfo {year}
		{2015})}\BibitemShut {NoStop}%
	\bibitem [{\citenamefont {Y\'ang}\ \emph {et~al.}(2020)\citenamefont {Y\'ang},
		\citenamefont {Cochran}, \citenamefont {Chapai}, \citenamefont {Tristant},
		\citenamefont {Yin}, \citenamefont {Belopolski}, \citenamefont {Ch\'eng},
		\citenamefont {Multer}, \citenamefont {Zhang}, \citenamefont {Shumiya},
		\citenamefont {Litskevich}, \citenamefont {Jiang}, \citenamefont {Chang},
		\citenamefont {Zhang}, \citenamefont {Vekhter}, \citenamefont {Shelton},
		\citenamefont {Jin}, \citenamefont {Xu},\ and\ \citenamefont
		{Hasan}}]{yang2020observation}%
	\BibitemOpen
	\bibfield  {author} {\bibinfo {author} {\bibfnamefont {X.}~\bibnamefont
			{Y\'ang}}, \bibinfo {author} {\bibfnamefont {T.~A.}\ \bibnamefont {Cochran}},
		\bibinfo {author} {\bibfnamefont {R.}~\bibnamefont {Chapai}}, \bibinfo
		{author} {\bibfnamefont {D.}~\bibnamefont {Tristant}}, \bibinfo {author}
		{\bibfnamefont {J.-X.}\ \bibnamefont {Yin}}, \bibinfo {author} {\bibfnamefont
			{I.}~\bibnamefont {Belopolski}}, \bibinfo {author} {\bibfnamefont {Z.~b. u.
				b.~a.}\ \bibnamefont {Ch\'eng}}, \bibinfo {author} {\bibfnamefont
			{D.}~\bibnamefont {Multer}}, \bibinfo {author} {\bibfnamefont {S.~S.}\
			\bibnamefont {Zhang}}, \bibinfo {author} {\bibfnamefont {N.}~\bibnamefont
			{Shumiya}}, \bibinfo {author} {\bibfnamefont {M.}~\bibnamefont {Litskevich}},
		\bibinfo {author} {\bibfnamefont {Y.}~\bibnamefont {Jiang}}, \bibinfo
		{author} {\bibfnamefont {G.}~\bibnamefont {Chang}}, \bibinfo {author}
		{\bibfnamefont {Q.}~\bibnamefont {Zhang}}, \bibinfo {author} {\bibfnamefont
			{I.}~\bibnamefont {Vekhter}}, \bibinfo {author} {\bibfnamefont {W.~A.}\
			\bibnamefont {Shelton}}, \bibinfo {author} {\bibfnamefont {R.}~\bibnamefont
			{Jin}}, \bibinfo {author} {\bibfnamefont {S.-Y.}\ \bibnamefont {Xu}},\ and\
		\bibinfo {author} {\bibfnamefont {M.~Z.}\ \bibnamefont {Hasan}},\ }\href
	{https://doi.org/10.1103/PhysRevB.101.201105} {\bibfield  {journal} {\bibinfo
			{journal} {Phys. Rev. B}\ }\textbf {\bibinfo {volume} {101}},\ \bibinfo
		{pages} {201105} (\bibinfo {year} {2020})}\BibitemShut {NoStop}%
	\bibitem [{\citenamefont {Thirupathaiah}\ \emph {et~al.}(2021)\citenamefont
		{Thirupathaiah}, \citenamefont {Kushnirenk}, \citenamefont {Koepernik},
		\citenamefont {Piening}, \citenamefont {Buechner}, \citenamefont {Aswartham},
		\citenamefont {van~den Brink}, \citenamefont {Borisenko},\ and\ \citenamefont
		{Fulga}}]{thirupathaiah2021sixfold}%
	\BibitemOpen
	\bibfield  {author} {\bibinfo {author} {\bibfnamefont {S.}~\bibnamefont
			{Thirupathaiah}}, \bibinfo {author} {\bibfnamefont {Y.}~\bibnamefont
			{Kushnirenk}}, \bibinfo {author} {\bibfnamefont {K.}~\bibnamefont
			{Koepernik}}, \bibinfo {author} {\bibfnamefont {B.~R.}\ \bibnamefont
			{Piening}}, \bibinfo {author} {\bibfnamefont {B.}~\bibnamefont {Buechner}},
		\bibinfo {author} {\bibfnamefont {S.}~\bibnamefont {Aswartham}}, \bibinfo
		{author} {\bibfnamefont {J.}~\bibnamefont {van~den Brink}}, \bibinfo {author}
		{\bibfnamefont {S.}~\bibnamefont {Borisenko}},\ and\ \bibinfo {author}
		{\bibfnamefont {I.~C.}\ \bibnamefont {Fulga}},\ }\href@noop {} {\bibfield
		{journal} {\bibinfo  {journal} {SciPost Phys.}\ }\textbf {\bibinfo {volume}
			{10}},\ \bibinfo {pages} {004} (\bibinfo {year} {2021})}\BibitemShut
	{NoStop}%
	\bibitem [{\citenamefont {Yang}\ \emph {et~al.}(2019)\citenamefont {Yang},
		\citenamefont {Sun}, \citenamefont {Xia}, \citenamefont {Xue}, \citenamefont
		{Gao}, \citenamefont {Ge}, \citenamefont {Jia}, \citenamefont {Yuan},
		\citenamefont {Chong},\ and\ \citenamefont {Zhang}}]{yang2019topological}%
	\BibitemOpen
	\bibfield  {author} {\bibinfo {author} {\bibfnamefont {Y.}~\bibnamefont
			{Yang}}, \bibinfo {author} {\bibfnamefont {H.-x.}\ \bibnamefont {Sun}},
		\bibinfo {author} {\bibfnamefont {J.-p.}\ \bibnamefont {Xia}}, \bibinfo
		{author} {\bibfnamefont {H.}~\bibnamefont {Xue}}, \bibinfo {author}
		{\bibfnamefont {Z.}~\bibnamefont {Gao}}, \bibinfo {author} {\bibfnamefont
			{Y.}~\bibnamefont {Ge}}, \bibinfo {author} {\bibfnamefont {D.}~\bibnamefont
			{Jia}}, \bibinfo {author} {\bibfnamefont {S.-q.}\ \bibnamefont {Yuan}},
		\bibinfo {author} {\bibfnamefont {Y.}~\bibnamefont {Chong}},\ and\ \bibinfo
		{author} {\bibfnamefont {B.}~\bibnamefont {Zhang}},\ }\href@noop {}
	{\bibfield  {journal} {\bibinfo  {journal} {Nat. Phys.}\ }\textbf {\bibinfo
			{volume} {15}},\ \bibinfo {pages} {645} (\bibinfo {year} {2019})}\BibitemShut
	{NoStop}%
	\bibitem [{\citenamefont {Hasan}\ \emph {et~al.}(2021)\citenamefont {Hasan},
		\citenamefont {Chang}, \citenamefont {Belopolski}, \citenamefont {Bian},
		\citenamefont {Xu},\ and\ \citenamefont {Yin}}]{hasan2021weyl}%
	\BibitemOpen
	\bibfield  {author} {\bibinfo {author} {\bibfnamefont {M.~Z.}\ \bibnamefont
			{Hasan}}, \bibinfo {author} {\bibfnamefont {G.}~\bibnamefont {Chang}},
		\bibinfo {author} {\bibfnamefont {I.}~\bibnamefont {Belopolski}}, \bibinfo
		{author} {\bibfnamefont {G.}~\bibnamefont {Bian}}, \bibinfo {author}
		{\bibfnamefont {S.-Y.}\ \bibnamefont {Xu}},\ and\ \bibinfo {author}
		{\bibfnamefont {J.-X.}\ \bibnamefont {Yin}},\ }\href@noop {} {\bibfield
		{journal} {\bibinfo  {journal} {Nat. Rev. Mater.}\ }\textbf {\bibinfo
			{volume} {6}},\ \bibinfo {pages} {784} (\bibinfo {year} {2021})}\BibitemShut
	{NoStop}%
	\bibitem [{\citenamefont {Lv}\ \emph {et~al.}(2017)\citenamefont {Lv},
		\citenamefont {Feng}, \citenamefont {Xu}, \citenamefont {Gao}, \citenamefont
		{Ma}, \citenamefont {Kong}, \citenamefont {Richard}, \citenamefont {Huang},
		\citenamefont {Strocov}, \citenamefont {Fang} \emph
		{et~al.}}]{lv2017observation}%
	\BibitemOpen
	\bibfield  {author} {\bibinfo {author} {\bibfnamefont {B.}~\bibnamefont
			{Lv}}, \bibinfo {author} {\bibfnamefont {Z.-L.}\ \bibnamefont {Feng}},
		\bibinfo {author} {\bibfnamefont {Q.-N.}\ \bibnamefont {Xu}}, \bibinfo
		{author} {\bibfnamefont {X.}~\bibnamefont {Gao}}, \bibinfo {author}
		{\bibfnamefont {J.-Z.}\ \bibnamefont {Ma}}, \bibinfo {author} {\bibfnamefont
			{L.-Y.}\ \bibnamefont {Kong}}, \bibinfo {author} {\bibfnamefont
			{P.}~\bibnamefont {Richard}}, \bibinfo {author} {\bibfnamefont {Y.-B.}\
			\bibnamefont {Huang}}, \bibinfo {author} {\bibfnamefont {V.}~\bibnamefont
			{Strocov}}, \bibinfo {author} {\bibfnamefont {C.}~\bibnamefont {Fang}}, \emph
		{et~al.},\ }\href@noop {} {\bibfield  {journal} {\bibinfo  {journal}
			{Nature}\ }\textbf {\bibinfo {volume} {546}},\ \bibinfo {pages} {627}
		(\bibinfo {year} {2017})}\BibitemShut {NoStop}%
	\bibitem [{\citenamefont {Kumar}\ \emph {et~al.}(2019)\citenamefont {Kumar},
		\citenamefont {Sun}, \citenamefont {Nicklas}, \citenamefont {Watzman},
		\citenamefont {Young}, \citenamefont {Leermakers}, \citenamefont {Hornung},
		\citenamefont {Klotz}, \citenamefont {Gooth}, \citenamefont {Manna} \emph
		{et~al.}}]{kumar2019extremely}%
	\BibitemOpen
	\bibfield  {author} {\bibinfo {author} {\bibfnamefont {N.}~\bibnamefont
			{Kumar}}, \bibinfo {author} {\bibfnamefont {Y.}~\bibnamefont {Sun}}, \bibinfo
		{author} {\bibfnamefont {M.}~\bibnamefont {Nicklas}}, \bibinfo {author}
		{\bibfnamefont {S.~J.}\ \bibnamefont {Watzman}}, \bibinfo {author}
		{\bibfnamefont {O.}~\bibnamefont {Young}}, \bibinfo {author} {\bibfnamefont
			{I.}~\bibnamefont {Leermakers}}, \bibinfo {author} {\bibfnamefont
			{J.}~\bibnamefont {Hornung}}, \bibinfo {author} {\bibfnamefont
			{J.}~\bibnamefont {Klotz}}, \bibinfo {author} {\bibfnamefont
			{J.}~\bibnamefont {Gooth}}, \bibinfo {author} {\bibfnamefont
			{K.}~\bibnamefont {Manna}}, \emph {et~al.},\ }\href@noop {} {\bibfield
		{journal} {\bibinfo  {journal} {Nat. Commun.}\ }\textbf {\bibinfo {volume}
			{10}},\ \bibinfo {pages} {1} (\bibinfo {year} {2019})}\BibitemShut {NoStop}%
	\bibitem [{\citenamefont {Zhu}\ \emph {et~al.}(2016)\citenamefont {Zhu},
		\citenamefont {Winkler}, \citenamefont {Wu}, \citenamefont {Li},\ and\
		\citenamefont {Soluyanov}}]{zhu2016triple}%
	\BibitemOpen
	\bibfield  {author} {\bibinfo {author} {\bibfnamefont {Z.}~\bibnamefont
			{Zhu}}, \bibinfo {author} {\bibfnamefont {G.~W.}\ \bibnamefont {Winkler}},
		\bibinfo {author} {\bibfnamefont {Q.}~\bibnamefont {Wu}}, \bibinfo {author}
		{\bibfnamefont {J.}~\bibnamefont {Li}},\ and\ \bibinfo {author}
		{\bibfnamefont {A.~A.}\ \bibnamefont {Soluyanov}},\ }\href@noop {} {\bibfield
		{journal} {\bibinfo  {journal} {Phys. Rev. X}\ }\textbf {\bibinfo {volume}
			{6}},\ \bibinfo {pages} {031003} (\bibinfo {year} {2016})}\BibitemShut
	{NoStop}%
	\bibitem [{\citenamefont {Ma}\ \emph {et~al.}(2018)\citenamefont {Ma},
		\citenamefont {He}, \citenamefont {Xu}, \citenamefont {Lv}, \citenamefont
		{Chen}, \citenamefont {Zhu}, \citenamefont {Zhang}, \citenamefont {Kong},
		\citenamefont {Gao}, \citenamefont {Rong} \emph {et~al.}}]{ma2018three}%
	\BibitemOpen
	\bibfield  {author} {\bibinfo {author} {\bibfnamefont {J.-Z.}\ \bibnamefont
			{Ma}}, \bibinfo {author} {\bibfnamefont {J.-B.}\ \bibnamefont {He}}, \bibinfo
		{author} {\bibfnamefont {Y.-F.}\ \bibnamefont {Xu}}, \bibinfo {author}
		{\bibfnamefont {B.}~\bibnamefont {Lv}}, \bibinfo {author} {\bibfnamefont
			{D.}~\bibnamefont {Chen}}, \bibinfo {author} {\bibfnamefont {W.-L.}\
			\bibnamefont {Zhu}}, \bibinfo {author} {\bibfnamefont {S.}~\bibnamefont
			{Zhang}}, \bibinfo {author} {\bibfnamefont {L.-Y.}\ \bibnamefont {Kong}},
		\bibinfo {author} {\bibfnamefont {X.}~\bibnamefont {Gao}}, \bibinfo {author}
		{\bibfnamefont {L.-Y.}\ \bibnamefont {Rong}}, \emph {et~al.},\ }\href@noop {}
	{\bibfield  {journal} {\bibinfo  {journal} {Nat. Phys.}\ }\textbf {\bibinfo
			{volume} {14}},\ \bibinfo {pages} {349} (\bibinfo {year} {2018})}\BibitemShut
	{NoStop}%
	\bibitem [{\citenamefont {Weng}\ \emph {et~al.}(2016)\citenamefont {Weng},
		\citenamefont {Fang}, \citenamefont {Fang},\ and\ \citenamefont
		{Dai}}]{weng2016topological}%
	\BibitemOpen
	\bibfield  {author} {\bibinfo {author} {\bibfnamefont {H.}~\bibnamefont
			{Weng}}, \bibinfo {author} {\bibfnamefont {C.}~\bibnamefont {Fang}}, \bibinfo
		{author} {\bibfnamefont {Z.}~\bibnamefont {Fang}},\ and\ \bibinfo {author}
		{\bibfnamefont {X.}~\bibnamefont {Dai}},\ }\href@noop {} {\bibfield
		{journal} {\bibinfo  {journal} {Phys. Rev. B}\ }\textbf {\bibinfo {volume}
			{93}},\ \bibinfo {pages} {241202} (\bibinfo {year} {2016})}\BibitemShut
	{NoStop}%
	\bibitem [{\citenamefont {Mardanya}\ \emph {et~al.}(2019)\citenamefont
		{Mardanya}, \citenamefont {Singh}, \citenamefont {Huang}, \citenamefont
		{Chang}, \citenamefont {Su}, \citenamefont {Lin}, \citenamefont {Agarwal},\
		and\ \citenamefont {Bansil}}]{Tripple_BaAgAs}%
	\BibitemOpen
	\bibfield  {author} {\bibinfo {author} {\bibfnamefont {S.}~\bibnamefont
			{Mardanya}}, \bibinfo {author} {\bibfnamefont {B.}~\bibnamefont {Singh}},
		\bibinfo {author} {\bibfnamefont {S.-M.}\ \bibnamefont {Huang}}, \bibinfo
		{author} {\bibfnamefont {T.-R.}\ \bibnamefont {Chang}}, \bibinfo {author}
		{\bibfnamefont {C.}~\bibnamefont {Su}}, \bibinfo {author} {\bibfnamefont
			{H.}~\bibnamefont {Lin}}, \bibinfo {author} {\bibfnamefont {A.}~\bibnamefont
			{Agarwal}},\ and\ \bibinfo {author} {\bibfnamefont {A.}~\bibnamefont
			{Bansil}},\ }\href {https://doi.org/10.1103/PhysRevMaterials.3.071201}
	{\bibfield  {journal} {\bibinfo  {journal} {Phys. Rev. Mater.}\ }\textbf
		{\bibinfo {volume} {3}},\ \bibinfo {pages} {071201} (\bibinfo {year}
		{2019})}\BibitemShut {NoStop}%
	\bibitem [{\citenamefont {Yan}\ \emph {et~al.}(2017)\citenamefont {Yan},
		\citenamefont {Bi}, \citenamefont {Shen}, \citenamefont {Lu}, \citenamefont
		{Zhang},\ and\ \citenamefont {Wang}}]{yan2017nodal}%
	\BibitemOpen
	\bibfield  {author} {\bibinfo {author} {\bibfnamefont {Z.}~\bibnamefont
			{Yan}}, \bibinfo {author} {\bibfnamefont {R.}~\bibnamefont {Bi}}, \bibinfo
		{author} {\bibfnamefont {H.}~\bibnamefont {Shen}}, \bibinfo {author}
		{\bibfnamefont {L.}~\bibnamefont {Lu}}, \bibinfo {author} {\bibfnamefont
			{S.-C.}\ \bibnamefont {Zhang}},\ and\ \bibinfo {author} {\bibfnamefont
			{Z.}~\bibnamefont {Wang}},\ }\href@noop {} {\bibfield  {journal} {\bibinfo
			{journal} {Phys. Rev. B}\ }\textbf {\bibinfo {volume} {96}},\ \bibinfo
		{pages} {041103} (\bibinfo {year} {2017})}\BibitemShut {NoStop}%
	\bibitem [{\citenamefont {Bi}\ \emph {et~al.}(2017)\citenamefont {Bi},
		\citenamefont {Yan}, \citenamefont {Lu},\ and\ \citenamefont
		{Wang}}]{bi2017nodal}%
	\BibitemOpen
	\bibfield  {author} {\bibinfo {author} {\bibfnamefont {R.}~\bibnamefont
			{Bi}}, \bibinfo {author} {\bibfnamefont {Z.}~\bibnamefont {Yan}}, \bibinfo
		{author} {\bibfnamefont {L.}~\bibnamefont {Lu}},\ and\ \bibinfo {author}
		{\bibfnamefont {Z.}~\bibnamefont {Wang}},\ }\href@noop {} {\bibfield
		{journal} {\bibinfo  {journal} {Phys. Rev. B}\ }\textbf {\bibinfo {volume}
			{96}},\ \bibinfo {pages} {201305} (\bibinfo {year} {2017})}\BibitemShut
	{NoStop}%
	\bibitem [{\citenamefont {Fang}\ \emph {et~al.}(2016)\citenamefont {Fang},
		\citenamefont {Weng}, \citenamefont {Dai},\ and\ \citenamefont
		{Fang}}]{fang2016topological}%
	\BibitemOpen
	\bibfield  {author} {\bibinfo {author} {\bibfnamefont {C.}~\bibnamefont
			{Fang}}, \bibinfo {author} {\bibfnamefont {H.}~\bibnamefont {Weng}}, \bibinfo
		{author} {\bibfnamefont {X.}~\bibnamefont {Dai}},\ and\ \bibinfo {author}
		{\bibfnamefont {Z.}~\bibnamefont {Fang}},\ }\href@noop {} {\bibfield
		{journal} {\bibinfo  {journal} {Chin. Phys. B}\ }\textbf {\bibinfo {volume}
			{25}},\ \bibinfo {pages} {117106} (\bibinfo {year} {2016})}\BibitemShut
	{NoStop}%
	\bibitem [{\citenamefont {Rui}\ \emph {et~al.}(2018)\citenamefont {Rui},
		\citenamefont {Zhao},\ and\ \citenamefont {Schnyder}}]{rui2018topological}%
	\BibitemOpen
	\bibfield  {author} {\bibinfo {author} {\bibfnamefont {W.}~\bibnamefont
			{Rui}}, \bibinfo {author} {\bibfnamefont {Y.}~\bibnamefont {Zhao}},\ and\
		\bibinfo {author} {\bibfnamefont {A.~P.}\ \bibnamefont {Schnyder}},\
	}\href@noop {} {\bibfield  {journal} {\bibinfo  {journal} {Phys. Rev. B}\
		}\textbf {\bibinfo {volume} {97}},\ \bibinfo {pages} {161113} (\bibinfo
		{year} {2018})}\BibitemShut {NoStop}%
	\bibitem [{\citenamefont {Zhu}\ \emph {et~al.}(2022)\citenamefont {Zhu},
		\citenamefont {Wu}, \citenamefont {Zhao}, \citenamefont {Chen}, \citenamefont
		{Zhang},\ and\ \citenamefont {Yang}}]{zhu2022symmetry}%
	\BibitemOpen
	\bibfield  {author} {\bibinfo {author} {\bibfnamefont {J.}~\bibnamefont
			{Zhu}}, \bibinfo {author} {\bibfnamefont {W.}~\bibnamefont {Wu}}, \bibinfo
		{author} {\bibfnamefont {J.}~\bibnamefont {Zhao}}, \bibinfo {author}
		{\bibfnamefont {H.}~\bibnamefont {Chen}}, \bibinfo {author} {\bibfnamefont
			{L.}~\bibnamefont {Zhang}},\ and\ \bibinfo {author} {\bibfnamefont {S.~A.}\
			\bibnamefont {Yang}},\ }\href@noop {} {\bibfield  {journal} {\bibinfo
			{journal} {npj Quantum Mater.}\ }\textbf {\bibinfo {volume} {7}},\ \bibinfo
		{pages} {1} (\bibinfo {year} {2022})}\BibitemShut {NoStop}%
	\bibitem [{\citenamefont {Zhang}\ \emph {et~al.}(2017)\citenamefont {Zhang},
		\citenamefont {Yu}, \citenamefont {Guo}, \citenamefont {Shi}, \citenamefont
		{Zhang},\ and\ \citenamefont {Yao}}]{zhang2017type}%
	\BibitemOpen
	\bibfield  {author} {\bibinfo {author} {\bibfnamefont {T.-T.}\ \bibnamefont
			{Zhang}}, \bibinfo {author} {\bibfnamefont {Z.-M.}\ \bibnamefont {Yu}},
		\bibinfo {author} {\bibfnamefont {W.}~\bibnamefont {Guo}}, \bibinfo {author}
		{\bibfnamefont {D.}~\bibnamefont {Shi}}, \bibinfo {author} {\bibfnamefont
			{G.}~\bibnamefont {Zhang}},\ and\ \bibinfo {author} {\bibfnamefont
			{Y.}~\bibnamefont {Yao}},\ }\href@noop {} {\bibfield  {journal} {\bibinfo
			{journal} {J. Phys. Chem. Lett.}\ }\textbf {\bibinfo {volume} {8}},\ \bibinfo
		{pages} {5792} (\bibinfo {year} {2017})}\BibitemShut {NoStop}%
	\bibitem [{\citenamefont {Soluyanov}(2017)}]{soluyanov2017type}%
	\BibitemOpen
	\bibfield  {author} {\bibinfo {author} {\bibfnamefont {A.~A.}\ \bibnamefont
			{Soluyanov}},\ }\href@noop {} {\bibfield  {journal} {\bibinfo  {journal}
			{Phys.}\ }\textbf {\bibinfo {volume} {10}},\ \bibinfo {pages} {74} (\bibinfo
		{year} {2017})}\BibitemShut {NoStop}%
	\bibitem [{\citenamefont {Sun}\ \emph {et~al.}(2015{\natexlab{b}})\citenamefont
		{Sun}, \citenamefont {Wu}, \citenamefont {Ali}, \citenamefont {Felser},\ and\
		\citenamefont {Yan}}]{sun2015prediction}%
	\BibitemOpen
	\bibfield  {author} {\bibinfo {author} {\bibfnamefont {Y.}~\bibnamefont
			{Sun}}, \bibinfo {author} {\bibfnamefont {S.-C.}\ \bibnamefont {Wu}},
		\bibinfo {author} {\bibfnamefont {M.~N.}\ \bibnamefont {Ali}}, \bibinfo
		{author} {\bibfnamefont {C.}~\bibnamefont {Felser}},\ and\ \bibinfo {author}
		{\bibfnamefont {B.}~\bibnamefont {Yan}},\ }\href@noop {} {\bibfield
		{journal} {\bibinfo  {journal} {Phys. Rev. B}\ }\textbf {\bibinfo {volume}
			{92}},\ \bibinfo {pages} {161107} (\bibinfo {year}
		{2015}{\natexlab{b}})}\BibitemShut {NoStop}%
	\bibitem [{\citenamefont {Xia}\ \emph {et~al.}(2019)\citenamefont {Xia},
		\citenamefont {Wang}, \citenamefont {Chen}, \citenamefont {Zhao},\ and\
		\citenamefont {Xu}}]{xia2019symmetry}%
	\BibitemOpen
	\bibfield  {author} {\bibinfo {author} {\bibfnamefont {B.}~\bibnamefont
			{Xia}}, \bibinfo {author} {\bibfnamefont {R.}~\bibnamefont {Wang}}, \bibinfo
		{author} {\bibfnamefont {Z.}~\bibnamefont {Chen}}, \bibinfo {author}
		{\bibfnamefont {Y.}~\bibnamefont {Zhao}},\ and\ \bibinfo {author}
		{\bibfnamefont {H.}~\bibnamefont {Xu}},\ }\href@noop {} {\bibfield  {journal}
		{\bibinfo  {journal} {Phys. Rev. Lett.}\ }\textbf {\bibinfo {volume} {123}},\
		\bibinfo {pages} {065501} (\bibinfo {year} {2019})}\BibitemShut {NoStop}%
	\bibitem [{\citenamefont {Xu}\ \emph {et~al.}(2017)\citenamefont {Xu},
		\citenamefont {Alidoust}, \citenamefont {Chang}, \citenamefont {Lu},
		\citenamefont {Singh}, \citenamefont {Belopolski}, \citenamefont {Sanchez},
		\citenamefont {Zhang}, \citenamefont {Bian}, \citenamefont {Zheng},
		\citenamefont {Husanu}, \citenamefont {Bian}, \citenamefont {Huang},
		\citenamefont {Hsu}, \citenamefont {Chang}, \citenamefont {Jeng},
		\citenamefont {Bansil}, \citenamefont {Neupert}, \citenamefont {Strocov},
		\citenamefont {Lin}, \citenamefont {Jia},\ and\ \citenamefont
		{Hasan}}]{LaAlGe_typeII}%
	\BibitemOpen
	\bibfield  {author} {\bibinfo {author} {\bibfnamefont {S.-Y.}\ \bibnamefont
			{Xu}}, \bibinfo {author} {\bibfnamefont {N.}~\bibnamefont {Alidoust}},
		\bibinfo {author} {\bibfnamefont {G.}~\bibnamefont {Chang}}, \bibinfo
		{author} {\bibfnamefont {H.}~\bibnamefont {Lu}}, \bibinfo {author}
		{\bibfnamefont {B.}~\bibnamefont {Singh}}, \bibinfo {author} {\bibfnamefont
			{I.}~\bibnamefont {Belopolski}}, \bibinfo {author} {\bibfnamefont {D.~S.}\
			\bibnamefont {Sanchez}}, \bibinfo {author} {\bibfnamefont {X.}~\bibnamefont
			{Zhang}}, \bibinfo {author} {\bibfnamefont {G.}~\bibnamefont {Bian}},
		\bibinfo {author} {\bibfnamefont {H.}~\bibnamefont {Zheng}}, \bibinfo
		{author} {\bibfnamefont {M.-A.}\ \bibnamefont {Husanu}}, \bibinfo {author}
		{\bibfnamefont {Y.}~\bibnamefont {Bian}}, \bibinfo {author} {\bibfnamefont
			{S.-M.}\ \bibnamefont {Huang}}, \bibinfo {author} {\bibfnamefont {C.-H.}\
			\bibnamefont {Hsu}}, \bibinfo {author} {\bibfnamefont {T.-R.}\ \bibnamefont
			{Chang}}, \bibinfo {author} {\bibfnamefont {H.-T.}\ \bibnamefont {Jeng}},
		\bibinfo {author} {\bibfnamefont {A.}~\bibnamefont {Bansil}}, \bibinfo
		{author} {\bibfnamefont {T.}~\bibnamefont {Neupert}}, \bibinfo {author}
		{\bibfnamefont {V.~N.}\ \bibnamefont {Strocov}}, \bibinfo {author}
		{\bibfnamefont {H.}~\bibnamefont {Lin}}, \bibinfo {author} {\bibfnamefont
			{S.}~\bibnamefont {Jia}},\ and\ \bibinfo {author} {\bibfnamefont {M.~Z.}\
			\bibnamefont {Hasan}},\ }\href {https://doi.org/10.1126/sciadv.1603266}
	{\bibfield  {journal} {\bibinfo  {journal} {Sci. Adv.}\ }\textbf {\bibinfo
			{volume} {3}},\ \bibinfo {pages} {e1603266} (\bibinfo {year}
		{2017})}\BibitemShut {NoStop}%
	\bibitem [{\citenamefont {Zhu}\ \emph {et~al.}(2018)\citenamefont {Zhu},
		\citenamefont {Yi}, \citenamefont {Li}, \citenamefont {Xiao}, \citenamefont
		{Zhang}, \citenamefont {Yang}, \citenamefont {Kaindl}, \citenamefont {Li},
		\citenamefont {Wang},\ and\ \citenamefont {Zhang}}]{zhu2018observation}%
	\BibitemOpen
	\bibfield  {author} {\bibinfo {author} {\bibfnamefont {H.}~\bibnamefont
			{Zhu}}, \bibinfo {author} {\bibfnamefont {J.}~\bibnamefont {Yi}}, \bibinfo
		{author} {\bibfnamefont {M.-Y.}\ \bibnamefont {Li}}, \bibinfo {author}
		{\bibfnamefont {J.}~\bibnamefont {Xiao}}, \bibinfo {author} {\bibfnamefont
			{L.}~\bibnamefont {Zhang}}, \bibinfo {author} {\bibfnamefont {C.-W.}\
			\bibnamefont {Yang}}, \bibinfo {author} {\bibfnamefont {R.~A.}\ \bibnamefont
			{Kaindl}}, \bibinfo {author} {\bibfnamefont {L.-J.}\ \bibnamefont {Li}},
		\bibinfo {author} {\bibfnamefont {Y.}~\bibnamefont {Wang}},\ and\ \bibinfo
		{author} {\bibfnamefont {X.}~\bibnamefont {Zhang}},\ }\href@noop {}
	{\bibfield  {journal} {\bibinfo  {journal} {Science}\ }\textbf {\bibinfo
			{volume} {359}},\ \bibinfo {pages} {579} (\bibinfo {year}
		{2018})}\BibitemShut {NoStop}%
	\bibitem [{\citenamefont {Wang}\ \emph
		{et~al.}(2021{\natexlab{a}})\citenamefont {Wang}, \citenamefont {Yuan},
		\citenamefont {Kuang}, \citenamefont {Yang}, \citenamefont {Yu},
		\citenamefont {Zhang},\ and\ \citenamefont {Wang}}]{wang2021coexistence}%
	\BibitemOpen
	\bibfield  {author} {\bibinfo {author} {\bibfnamefont {J.}~\bibnamefont
			{Wang}}, \bibinfo {author} {\bibfnamefont {H.}~\bibnamefont {Yuan}}, \bibinfo
		{author} {\bibfnamefont {M.}~\bibnamefont {Kuang}}, \bibinfo {author}
		{\bibfnamefont {T.}~\bibnamefont {Yang}}, \bibinfo {author} {\bibfnamefont
			{Z.-M.}\ \bibnamefont {Yu}}, \bibinfo {author} {\bibfnamefont
			{Z.}~\bibnamefont {Zhang}},\ and\ \bibinfo {author} {\bibfnamefont
			{X.}~\bibnamefont {Wang}},\ }\href@noop {} {\bibfield  {journal} {\bibinfo
			{journal} {Phys. Rev. B}\ }\textbf {\bibinfo {volume} {104}},\ \bibinfo
		{pages} {L041107} (\bibinfo {year} {2021}{\natexlab{a}})}\BibitemShut
	{NoStop}%
	\bibitem [{\citenamefont {Litvinchuk}\ and\ \citenamefont
		{Valakh}(2020)}]{litvinchuk2020raman}%
	\BibitemOpen
	\bibfield  {author} {\bibinfo {author} {\bibfnamefont {A.~P.}\ \bibnamefont
			{Litvinchuk}}\ and\ \bibinfo {author} {\bibfnamefont {M.~Y.}\ \bibnamefont
			{Valakh}},\ }\href@noop {} {\bibfield  {journal} {\bibinfo  {journal} {J.
				Condens. Matter Phys.}\ }\textbf {\bibinfo {volume} {32}},\ \bibinfo {pages}
		{445401} (\bibinfo {year} {2020})}\BibitemShut {NoStop}%
	\bibitem [{\citenamefont {Li}\ \emph {et~al.}(2019)\citenamefont {Li},
		\citenamefont {Wang}, \citenamefont {Liu}, \citenamefont {Li}, \citenamefont
		{Zhang},\ and\ \citenamefont {Chen}}]{li2019topological}%
	\BibitemOpen
	\bibfield  {author} {\bibinfo {author} {\bibfnamefont {J.}~\bibnamefont
			{Li}}, \bibinfo {author} {\bibfnamefont {L.}~\bibnamefont {Wang}}, \bibinfo
		{author} {\bibfnamefont {J.}~\bibnamefont {Liu}}, \bibinfo {author}
		{\bibfnamefont {R.}~\bibnamefont {Li}}, \bibinfo {author} {\bibfnamefont
			{Z.}~\bibnamefont {Zhang}},\ and\ \bibinfo {author} {\bibfnamefont {X.-Q.}\
			\bibnamefont {Chen}},\ }\href@noop {} {\bibfield  {journal} {\bibinfo
			{journal} {arXiv preprint arXiv:1907.08547}\ } (\bibinfo {year}
		{2019})}\BibitemShut {NoStop}%
	\bibitem [{\citenamefont {Miao}\ \emph {et~al.}(2018)\citenamefont {Miao},
		\citenamefont {Zhang}, \citenamefont {Wang}, \citenamefont {Meyers},
		\citenamefont {Said}, \citenamefont {Wang}, \citenamefont {Shi},
		\citenamefont {Weng}, \citenamefont {Fang},\ and\ \citenamefont
		{Dean}}]{miao2018observation}%
	\BibitemOpen
	\bibfield  {author} {\bibinfo {author} {\bibfnamefont {H.}~\bibnamefont
			{Miao}}, \bibinfo {author} {\bibfnamefont {T.}~\bibnamefont {Zhang}},
		\bibinfo {author} {\bibfnamefont {L.}~\bibnamefont {Wang}}, \bibinfo {author}
		{\bibfnamefont {D.}~\bibnamefont {Meyers}}, \bibinfo {author} {\bibfnamefont
			{A.}~\bibnamefont {Said}}, \bibinfo {author} {\bibfnamefont {Y.}~\bibnamefont
			{Wang}}, \bibinfo {author} {\bibfnamefont {Y.}~\bibnamefont {Shi}}, \bibinfo
		{author} {\bibfnamefont {H.}~\bibnamefont {Weng}}, \bibinfo {author}
		{\bibfnamefont {Z.}~\bibnamefont {Fang}},\ and\ \bibinfo {author}
		{\bibfnamefont {M.}~\bibnamefont {Dean}},\ }\href@noop {} {\bibfield
		{journal} {\bibinfo  {journal} {Phys. Rev. Lett.}\ }\textbf {\bibinfo
			{volume} {121}},\ \bibinfo {pages} {035302} (\bibinfo {year}
		{2018})}\BibitemShut {NoStop}%
	\bibitem [{\citenamefont {Liu}\ \emph {et~al.}(2021)\citenamefont {Liu},
		\citenamefont {Wang},\ and\ \citenamefont {Fu}}]{liu2021charge}%
	\BibitemOpen
	\bibfield  {author} {\bibinfo {author} {\bibfnamefont {Q.-B.}\ \bibnamefont
			{Liu}}, \bibinfo {author} {\bibfnamefont {Z.}~\bibnamefont {Wang}},\ and\
		\bibinfo {author} {\bibfnamefont {H.-H.}\ \bibnamefont {Fu}},\ }\href@noop {}
	{\bibfield  {journal} {\bibinfo  {journal} {Phys. Rev. B}\ }\textbf {\bibinfo
			{volume} {103}},\ \bibinfo {pages} {L161303} (\bibinfo {year}
		{2021})}\BibitemShut {NoStop}%
	\bibitem [{\citenamefont {Wang}\ \emph
		{et~al.}(2021{\natexlab{b}})\citenamefont {Wang}, \citenamefont {Zhou},
		\citenamefont {Yang}, \citenamefont {Kuang}, \citenamefont {Yu},\ and\
		\citenamefont {Zhang}}]{wang2021symmetry}%
	\BibitemOpen
	\bibfield  {author} {\bibinfo {author} {\bibfnamefont {X.}~\bibnamefont
			{Wang}}, \bibinfo {author} {\bibfnamefont {F.}~\bibnamefont {Zhou}}, \bibinfo
		{author} {\bibfnamefont {T.}~\bibnamefont {Yang}}, \bibinfo {author}
		{\bibfnamefont {M.}~\bibnamefont {Kuang}}, \bibinfo {author} {\bibfnamefont
			{Z.-M.}\ \bibnamefont {Yu}},\ and\ \bibinfo {author} {\bibfnamefont
			{G.}~\bibnamefont {Zhang}},\ }\href@noop {} {\bibfield  {journal} {\bibinfo
			{journal} {Phys. Rev. B}\ }\textbf {\bibinfo {volume} {104}},\ \bibinfo
		{pages} {L041104} (\bibinfo {year} {2021}{\natexlab{b}})}\BibitemShut
	{NoStop}%
	\bibitem [{\citenamefont {Xie}\ \emph {et~al.}(2021)\citenamefont {Xie},
		\citenamefont {Liu}, \citenamefont {Zhang}, \citenamefont {Zhou},
		\citenamefont {Yang}, \citenamefont {Kuang}, \citenamefont {Wang},\ and\
		\citenamefont {Zhang}}]{xie2021sixfold}%
	\BibitemOpen
	\bibfield  {author} {\bibinfo {author} {\bibfnamefont {C.}~\bibnamefont
			{Xie}}, \bibinfo {author} {\bibfnamefont {Y.}~\bibnamefont {Liu}}, \bibinfo
		{author} {\bibfnamefont {Z.}~\bibnamefont {Zhang}}, \bibinfo {author}
		{\bibfnamefont {F.}~\bibnamefont {Zhou}}, \bibinfo {author} {\bibfnamefont
			{T.}~\bibnamefont {Yang}}, \bibinfo {author} {\bibfnamefont {M.}~\bibnamefont
			{Kuang}}, \bibinfo {author} {\bibfnamefont {X.}~\bibnamefont {Wang}},\ and\
		\bibinfo {author} {\bibfnamefont {G.}~\bibnamefont {Zhang}},\ }\href@noop {}
	{\bibfield  {journal} {\bibinfo  {journal} {Phys. Rev. B}\ }\textbf {\bibinfo
			{volume} {104}},\ \bibinfo {pages} {045148} (\bibinfo {year}
		{2021})}\BibitemShut {NoStop}%
	\bibitem [{\citenamefont {Liu}\ \emph {et~al.}(2020)\citenamefont {Liu},
		\citenamefont {Chen},\ and\ \citenamefont {Xu}}]{liu2020topological}%
	\BibitemOpen
	\bibfield  {author} {\bibinfo {author} {\bibfnamefont {Y.}~\bibnamefont
			{Liu}}, \bibinfo {author} {\bibfnamefont {X.}~\bibnamefont {Chen}},\ and\
		\bibinfo {author} {\bibfnamefont {Y.}~\bibnamefont {Xu}},\ }\href@noop {}
	{\bibfield  {journal} {\bibinfo  {journal} {Adv. Funct. Mater.}\ }\textbf
		{\bibinfo {volume} {30}},\ \bibinfo {pages} {1904784} (\bibinfo {year}
		{2020})}\BibitemShut {NoStop}%
	\bibitem [{\citenamefont {Zhong}\ \emph {et~al.}(2021)\citenamefont {Zhong},
		\citenamefont {Liu}, \citenamefont {Zhou}, \citenamefont {Kuang},
		\citenamefont {Yang}, \citenamefont {Wang},\ and\ \citenamefont
		{Zhang}}]{zhong2021coexistence}%
	\BibitemOpen
	\bibfield  {author} {\bibinfo {author} {\bibfnamefont {M.}~\bibnamefont
			{Zhong}}, \bibinfo {author} {\bibfnamefont {Y.}~\bibnamefont {Liu}}, \bibinfo
		{author} {\bibfnamefont {F.}~\bibnamefont {Zhou}}, \bibinfo {author}
		{\bibfnamefont {M.}~\bibnamefont {Kuang}}, \bibinfo {author} {\bibfnamefont
			{T.}~\bibnamefont {Yang}}, \bibinfo {author} {\bibfnamefont {X.}~\bibnamefont
			{Wang}},\ and\ \bibinfo {author} {\bibfnamefont {G.}~\bibnamefont {Zhang}},\
	}\href@noop {} {\bibfield  {journal} {\bibinfo  {journal} {Phys. Rev. B}\
		}\textbf {\bibinfo {volume} {104}},\ \bibinfo {pages} {085118} (\bibinfo
		{year} {2021})}\BibitemShut {NoStop}%
	\bibitem [{\citenamefont {Perdew}\ \emph {et~al.}(1996)\citenamefont {Perdew},
		\citenamefont {Burke},\ and\ \citenamefont
		{Ernzerhof}}]{perdew1996generalized}%
	\BibitemOpen
	\bibfield  {author} {\bibinfo {author} {\bibfnamefont {J.~P.}\ \bibnamefont
			{Perdew}}, \bibinfo {author} {\bibfnamefont {K.}~\bibnamefont {Burke}},\ and\
		\bibinfo {author} {\bibfnamefont {M.}~\bibnamefont {Ernzerhof}},\ }\href@noop
	{} {\bibfield  {journal} {\bibinfo  {journal} {Phys. Rev. Lett.}\ }\textbf
		{\bibinfo {volume} {77}},\ \bibinfo {pages} {3865} (\bibinfo {year}
		{1996})}\BibitemShut {NoStop}%
	\bibitem [{\citenamefont {Tran}\ and\ \citenamefont
		{Blaha}(2009)}]{tran2009accurate}%
	\BibitemOpen
	\bibfield  {author} {\bibinfo {author} {\bibfnamefont {F.}~\bibnamefont
			{Tran}}\ and\ \bibinfo {author} {\bibfnamefont {P.}~\bibnamefont {Blaha}},\
	}\href@noop {} {\bibfield  {journal} {\bibinfo  {journal} {Phys. Rev. Lett.}\
		}\textbf {\bibinfo {volume} {102}},\ \bibinfo {pages} {226401} (\bibinfo
		{year} {2009})}\BibitemShut {NoStop}%
	\bibitem [{\citenamefont {McCrae}\ \emph {et~al.}(1997)\citenamefont {McCrae},
		\citenamefont {Hengehold}, \citenamefont {Yeo}, \citenamefont {Ohmer},\ and\
		\citenamefont {Schunemann}}]{mccrae1997photoluminescence}%
	\BibitemOpen
	\bibfield  {author} {\bibinfo {author} {\bibfnamefont {J.}~\bibnamefont
			{McCrae}}, \bibinfo {author} {\bibfnamefont {R.}~\bibnamefont {Hengehold}},
		\bibinfo {author} {\bibfnamefont {Y.}~\bibnamefont {Yeo}}, \bibinfo {author}
		{\bibfnamefont {M.}~\bibnamefont {Ohmer}},\ and\ \bibinfo {author}
		{\bibfnamefont {P.}~\bibnamefont {Schunemann}},\ }\href@noop {} {\bibfield
		{journal} {\bibinfo  {journal} {Appl. Phys. Lett.}\ }\textbf {\bibinfo
			{volume} {70}},\ \bibinfo {pages} {455} (\bibinfo {year} {1997})}\BibitemShut
	{NoStop}%
	\bibitem [{\citenamefont {Akimchenko}\ \emph {et~al.}(1973)\citenamefont
		{Akimchenko}, \citenamefont {Ivanov},\ and\ \citenamefont
		{Borshchevsky}}]{akimchenko1973electroreflection}%
	\BibitemOpen
	\bibfield  {author} {\bibinfo {author} {\bibfnamefont {I.}~\bibnamefont
			{Akimchenko}}, \bibinfo {author} {\bibfnamefont {V.}~\bibnamefont {Ivanov}},\
		and\ \bibinfo {author} {\bibfnamefont {A.}~\bibnamefont {Borshchevsky}},\
	}\href@noop {} {\bibfield  {journal} {\bibinfo  {journal} {Sov. Phys.
				Semiconduct.}\ }\textbf {\bibinfo {volume} {7}},\ \bibinfo {pages} {309}
		(\bibinfo {year} {1973})}\BibitemShut {NoStop}%
	\bibitem [{\citenamefont {Bai}\ \emph {et~al.}(2005)\citenamefont {Bai},
		\citenamefont {Xu}, \citenamefont {Schunemann}, \citenamefont {Nagashio},
		\citenamefont {Feigelson},\ and\ \citenamefont {Giles}}]{bai2005urbach}%
	\BibitemOpen
	\bibfield  {author} {\bibinfo {author} {\bibfnamefont {L.}~\bibnamefont
			{Bai}}, \bibinfo {author} {\bibfnamefont {C.}~\bibnamefont {Xu}}, \bibinfo
		{author} {\bibfnamefont {P.}~\bibnamefont {Schunemann}}, \bibinfo {author}
		{\bibfnamefont {K.}~\bibnamefont {Nagashio}}, \bibinfo {author}
		{\bibfnamefont {R.}~\bibnamefont {Feigelson}},\ and\ \bibinfo {author}
		{\bibfnamefont {N.}~\bibnamefont {Giles}},\ }\href@noop {} {\bibfield
		{journal} {\bibinfo  {journal} {J. Phys. Condens. Matter}\ }\textbf {\bibinfo
			{volume} {17}},\ \bibinfo {pages} {549} (\bibinfo {year} {2005})}\BibitemShut
	{NoStop}%
	\bibitem [{\citenamefont {Blaha}\ \emph {et~al.}(2001)\citenamefont {Blaha},
		\citenamefont {Schwarz}, \citenamefont {Madsen}, \citenamefont {Kvasnicka},
		\citenamefont {Luitz} \emph {et~al.}}]{blaha2001wien2k}%
	\BibitemOpen
	\bibfield  {author} {\bibinfo {author} {\bibfnamefont {P.}~\bibnamefont
			{Blaha}}, \bibinfo {author} {\bibfnamefont {K.}~\bibnamefont {Schwarz}},
		\bibinfo {author} {\bibfnamefont {G.~K.}\ \bibnamefont {Madsen}}, \bibinfo
		{author} {\bibfnamefont {D.}~\bibnamefont {Kvasnicka}}, \bibinfo {author}
		{\bibfnamefont {J.}~\bibnamefont {Luitz}}, \emph {et~al.},\ }\href@noop {}
	{\bibfield  {journal} {\bibinfo  {journal} {An augmented plane wave+ local
				orbitals program for calculating crystal properties}\ }\textbf {\bibinfo
			{volume} {60}} (\bibinfo {year} {2001})}\BibitemShut {NoStop}%
	\bibitem [{\citenamefont {Schwarz}\ and\ \citenamefont
		{Blaha}(2003)}]{schwarz2003solid}%
	\BibitemOpen
	\bibfield  {author} {\bibinfo {author} {\bibfnamefont {K.}~\bibnamefont
			{Schwarz}}\ and\ \bibinfo {author} {\bibfnamefont {P.}~\bibnamefont
			{Blaha}},\ }\href@noop {} {\bibfield  {journal} {\bibinfo  {journal} {Comput.
				Mater. Sci.}\ }\textbf {\bibinfo {volume} {28}},\ \bibinfo {pages} {259}
		(\bibinfo {year} {2003})}\BibitemShut {NoStop}%
	\bibitem [{\citenamefont {Schwarz}(2003)}]{schwarz2003dft}%
	\BibitemOpen
	\bibfield  {author} {\bibinfo {author} {\bibfnamefont {K.}~\bibnamefont
			{Schwarz}},\ }\href@noop {} {\bibfield  {journal} {\bibinfo  {journal} {J.
				Solid State Chem.}\ }\textbf {\bibinfo {volume} {176}},\ \bibinfo {pages}
		{319} (\bibinfo {year} {2003})}\BibitemShut {NoStop}%
	\bibitem [{\citenamefont {Madsen}\ and\ \citenamefont
		{Singh}(2006)}]{madsen2006boltztrap}%
	\BibitemOpen
	\bibfield  {author} {\bibinfo {author} {\bibfnamefont {G.~K.}\ \bibnamefont
			{Madsen}}\ and\ \bibinfo {author} {\bibfnamefont {D.~J.}\ \bibnamefont
			{Singh}},\ }\href@noop {} {\bibfield  {journal} {\bibinfo  {journal} {Comput.
				Phys. Commun.}\ }\textbf {\bibinfo {volume} {175}},\ \bibinfo {pages} {67}
		(\bibinfo {year} {2006})}\BibitemShut {NoStop}%
	\bibitem [{\citenamefont {Madsen}\ \emph {et~al.}(2018)\citenamefont {Madsen},
		\citenamefont {Carrete},\ and\ \citenamefont
		{Verstraete}}]{madsen2018boltztrap2}%
	\BibitemOpen
	\bibfield  {author} {\bibinfo {author} {\bibfnamefont {G.~K.}\ \bibnamefont
			{Madsen}}, \bibinfo {author} {\bibfnamefont {J.}~\bibnamefont {Carrete}},\
		and\ \bibinfo {author} {\bibfnamefont {M.~J.}\ \bibnamefont {Verstraete}},\
	}\href@noop {} {\bibfield  {journal} {\bibinfo  {journal} {Comput. Phys.
				Commun.}\ }\textbf {\bibinfo {volume} {231}},\ \bibinfo {pages} {140}
		(\bibinfo {year} {2018})}\BibitemShut {NoStop}%
	\bibitem [{\citenamefont {Kresse}\ and\ \citenamefont
		{Hafner}(1993)}]{kresse1993ab}%
	\BibitemOpen
	\bibfield  {author} {\bibinfo {author} {\bibfnamefont {G.}~\bibnamefont
			{Kresse}}\ and\ \bibinfo {author} {\bibfnamefont {J.}~\bibnamefont
			{Hafner}},\ }\href@noop {} {\bibfield  {journal} {\bibinfo  {journal} {Phys.
				Rev. B}\ }\textbf {\bibinfo {volume} {47}},\ \bibinfo {pages} {558} (\bibinfo
		{year} {1993})}\BibitemShut {NoStop}%
	\bibitem [{\citenamefont {Kresse}\ and\ \citenamefont
		{Furthm\"uller}(1996)}]{kresse1996efficient}%
	\BibitemOpen
	\bibfield  {author} {\bibinfo {author} {\bibfnamefont {G.}~\bibnamefont
			{Kresse}}\ and\ \bibinfo {author} {\bibfnamefont {J.}~\bibnamefont
			{Furthm\"uller}},\ }\href@noop {} {\bibfield  {journal} {\bibinfo  {journal}
			{Phys. Rev. B}\ }\textbf {\bibinfo {volume} {54}},\ \bibinfo {pages} {11169}
		(\bibinfo {year} {1996})}\BibitemShut {NoStop}%
	\bibitem [{\citenamefont {Kresse}\ and\ \citenamefont
		{Joubert}(1999)}]{kresse1999from}%
	\BibitemOpen
	\bibfield  {author} {\bibinfo {author} {\bibfnamefont {G.}~\bibnamefont
			{Kresse}}\ and\ \bibinfo {author} {\bibfnamefont {D.}~\bibnamefont
			{Joubert}},\ }\href@noop {} {\bibfield  {journal} {\bibinfo  {journal} {Phys.
				Rev. B}\ }\textbf {\bibinfo {volume} {59}},\ \bibinfo {pages} {1758}
		(\bibinfo {year} {1999})}\BibitemShut {NoStop}%
	\bibitem [{\citenamefont {Togo}\ and\ \citenamefont
		{Tanaka}(2015)}]{togo2015first}%
	\BibitemOpen
	\bibfield  {author} {\bibinfo {author} {\bibfnamefont {A.}~\bibnamefont
			{Togo}}\ and\ \bibinfo {author} {\bibfnamefont {I.}~\bibnamefont {Tanaka}},\
	}\href@noop {} {\bibfield  {journal} {\bibinfo  {journal} {Scr. Mater.}\
		}\textbf {\bibinfo {volume} {108}},\ \bibinfo {pages} {1} (\bibinfo {year}
		{2015})}\BibitemShut {NoStop}%
	\bibitem [{\citenamefont {Singh}\ \emph {et~al.}(2018)\citenamefont {Singh},
		\citenamefont {Wu}, \citenamefont {Yue}, \citenamefont {Romero},\ and\
		\citenamefont {Soluyanov}}]{singh2018topological}%
	\BibitemOpen
	\bibfield  {author} {\bibinfo {author} {\bibfnamefont {S.}~\bibnamefont
			{Singh}}, \bibinfo {author} {\bibfnamefont {Q.}~\bibnamefont {Wu}}, \bibinfo
		{author} {\bibfnamefont {C.}~\bibnamefont {Yue}}, \bibinfo {author}
		{\bibfnamefont {A.~H.}\ \bibnamefont {Romero}},\ and\ \bibinfo {author}
		{\bibfnamefont {A.~A.}\ \bibnamefont {Soluyanov}},\ }\href@noop {} {\bibfield
		{journal} {\bibinfo  {journal} {Phys. Rev. Mater.}\ }\textbf {\bibinfo
			{volume} {2}},\ \bibinfo {pages} {114204} (\bibinfo {year}
		{2018})}\BibitemShut {NoStop}%
	\bibitem [{\citenamefont {Li}\ \emph {et~al.}(2016)\citenamefont {Li},
		\citenamefont {Carrete}, \citenamefont {Madsen},\ and\ \citenamefont
		{Mingo}}]{li2016influence}%
	\BibitemOpen
	\bibfield  {author} {\bibinfo {author} {\bibfnamefont {W.}~\bibnamefont
			{Li}}, \bibinfo {author} {\bibfnamefont {J.}~\bibnamefont {Carrete}},
		\bibinfo {author} {\bibfnamefont {G.~K.}\ \bibnamefont {Madsen}},\ and\
		\bibinfo {author} {\bibfnamefont {N.}~\bibnamefont {Mingo}},\ }\href@noop {}
	{\bibfield  {journal} {\bibinfo  {journal} {Phys. Rev. B}\ }\textbf {\bibinfo
			{volume} {93}},\ \bibinfo {pages} {205203} (\bibinfo {year}
		{2016})}\BibitemShut {NoStop}%
	\bibitem [{\citenamefont {Yu}\ and\ \citenamefont
		{Hong}(2021)}]{yu2021absence}%
	\BibitemOpen
	\bibfield  {author} {\bibinfo {author} {\bibfnamefont {X.}~\bibnamefont
			{Yu}}\ and\ \bibinfo {author} {\bibfnamefont {J.}~\bibnamefont {Hong}},\
	}\href@noop {} {\bibfield  {journal} {\bibinfo  {journal} {J. Mater. Chem.}\
		}\textbf {\bibinfo {volume} {9}},\ \bibinfo {pages} {12420} (\bibinfo {year}
		{2021})}\BibitemShut {NoStop}%
	\bibitem [{\citenamefont {Singh}(2010{\natexlab{a}})}]{singh2010doping}%
	\BibitemOpen
	\bibfield  {author} {\bibinfo {author} {\bibfnamefont {D.~J.}\ \bibnamefont
			{Singh}},\ }\href@noop {} {\bibfield  {journal} {\bibinfo  {journal} {Phys.
				Rev. B}\ }\textbf {\bibinfo {volume} {81}},\ \bibinfo {pages} {195217}
		(\bibinfo {year} {2010}{\natexlab{a}})}\BibitemShut {NoStop}%
	\bibitem [{\citenamefont {Parker}\ and\ \citenamefont
		{Singh}(2010)}]{parker2010high}%
	\BibitemOpen
	\bibfield  {author} {\bibinfo {author} {\bibfnamefont {D.}~\bibnamefont
			{Parker}}\ and\ \bibinfo {author} {\bibfnamefont {D.~J.}\ \bibnamefont
			{Singh}},\ }\href@noop {} {\bibfield  {journal} {\bibinfo  {journal} {Phys.
				Rev. B}\ }\textbf {\bibinfo {volume} {82}},\ \bibinfo {pages} {035204}
		(\bibinfo {year} {2010})}\BibitemShut {NoStop}%
	\bibitem [{\citenamefont {Wang}\ \emph {et~al.}(2011)\citenamefont {Wang},
		\citenamefont {Pei}, \citenamefont {LaLonde},\ and\ \citenamefont
		{Snyder}}]{wang2011heavily}%
	\BibitemOpen
	\bibfield  {author} {\bibinfo {author} {\bibfnamefont {H.}~\bibnamefont
			{Wang}}, \bibinfo {author} {\bibfnamefont {Y.}~\bibnamefont {Pei}}, \bibinfo
		{author} {\bibfnamefont {A.~D.}\ \bibnamefont {LaLonde}},\ and\ \bibinfo
		{author} {\bibfnamefont {G.~J.}\ \bibnamefont {Snyder}},\ }\href@noop {}
	{\bibfield  {journal} {\bibinfo  {journal} {Adv. Mater.}\ }\textbf {\bibinfo
			{volume} {23}},\ \bibinfo {pages} {1366} (\bibinfo {year}
		{2011})}\BibitemShut {NoStop}%
	\bibitem [{\citenamefont {Sun}\ and\ \citenamefont
		{Singh}(2016{\natexlab{a}})}]{sun2016thermoelectric}%
	\BibitemOpen
	\bibfield  {author} {\bibinfo {author} {\bibfnamefont {J.}~\bibnamefont
			{Sun}}\ and\ \bibinfo {author} {\bibfnamefont {D.~J.}\ \bibnamefont
			{Singh}},\ }\href@noop {} {\bibfield  {journal} {\bibinfo  {journal} {Phys.
				Rev. Appl.}\ }\textbf {\bibinfo {volume} {5}},\ \bibinfo {pages} {024006}
		(\bibinfo {year} {2016}{\natexlab{a}})}\BibitemShut {NoStop}%
	\bibitem [{\citenamefont {Sun}\ and\ \citenamefont
		{Singh}(2016{\natexlab{b}})}]{sun2016ther}%
	\BibitemOpen
	\bibfield  {author} {\bibinfo {author} {\bibfnamefont {J.}~\bibnamefont
			{Sun}}\ and\ \bibinfo {author} {\bibfnamefont {D.~J.}\ \bibnamefont
			{Singh}},\ }\href@noop {} {\bibfield  {journal} {\bibinfo  {journal} {APL
				Mater.}\ }\textbf {\bibinfo {volume} {4}},\ \bibinfo {pages} {104803}
		(\bibinfo {year} {2016}{\natexlab{b}})}\BibitemShut {NoStop}%
	\bibitem [{\citenamefont {He}\ and\ \citenamefont
		{Tritt}(2017)}]{he2017advances}%
	\BibitemOpen
	\bibfield  {author} {\bibinfo {author} {\bibfnamefont {J.}~\bibnamefont
			{He}}\ and\ \bibinfo {author} {\bibfnamefont {T.~M.}\ \bibnamefont {Tritt}},\
	}\href@noop {} {\bibfield  {journal} {\bibinfo  {journal} {Science}\ }\textbf
		{\bibinfo {volume} {357}},\ \bibinfo {pages} {eaak9997} (\bibinfo {year}
		{2017})}\BibitemShut {NoStop}%
	\bibitem [{\citenamefont {Irkhin}\ and\ \citenamefont
		{Irkhin}(2007)}]{irkhin2007electronic}%
	\BibitemOpen
	\bibfield  {author} {\bibinfo {author} {\bibfnamefont {V.~Y.}\ \bibnamefont
			{Irkhin}}\ and\ \bibinfo {author} {\bibfnamefont {Y.~P.}\ \bibnamefont
			{Irkhin}},\ }\href@noop {} {}\ (\bibinfo  {publisher} {Cambridge Int Science
		Publishing},\ \bibinfo {year} {2007})\BibitemShut {NoStop}%
	\bibitem [{\citenamefont {Dehkordi}\ \emph {et~al.}(2015)\citenamefont
		{Dehkordi}, \citenamefont {Zebarjadi}, \citenamefont {He},\ and\
		\citenamefont {Tritt}}]{dehkordi2015thermoelectric}%
	\BibitemOpen
	\bibfield  {author} {\bibinfo {author} {\bibfnamefont {A.~M.}\ \bibnamefont
			{Dehkordi}}, \bibinfo {author} {\bibfnamefont {M.}~\bibnamefont {Zebarjadi}},
		\bibinfo {author} {\bibfnamefont {J.}~\bibnamefont {He}},\ and\ \bibinfo
		{author} {\bibfnamefont {T.~M.}\ \bibnamefont {Tritt}},\ }\href@noop {}
	{\bibfield  {journal} {\bibinfo  {journal} {Mater. Sci. Eng. R Rep.}\
		}\textbf {\bibinfo {volume} {97}},\ \bibinfo {pages} {1} (\bibinfo {year}
		{2015})}\BibitemShut {NoStop}%
	\bibitem [{\citenamefont {Ekuma}\ \emph {et~al.}(2012)\citenamefont {Ekuma},
		\citenamefont {Singh}, \citenamefont {Moreno},\ and\ \citenamefont
		{Jarrell}}]{ekuma2012optical}%
	\BibitemOpen
	\bibfield  {author} {\bibinfo {author} {\bibfnamefont {C.~E.}\ \bibnamefont
			{Ekuma}}, \bibinfo {author} {\bibfnamefont {D.~J.}\ \bibnamefont {Singh}},
		\bibinfo {author} {\bibfnamefont {J.}~\bibnamefont {Moreno}},\ and\ \bibinfo
		{author} {\bibfnamefont {M.}~\bibnamefont {Jarrell}},\ }\href@noop {}
	{\bibfield  {journal} {\bibinfo  {journal} {Phys. Rev. B}\ }\textbf {\bibinfo
			{volume} {85}},\ \bibinfo {pages} {085205} (\bibinfo {year}
		{2012})}\BibitemShut {NoStop}%
	\bibitem [{\citenamefont {Singh}(2010{\natexlab{b}})}]{singh2010thermopower}%
	\BibitemOpen
	\bibfield  {author} {\bibinfo {author} {\bibfnamefont {D.~J.}\ \bibnamefont
			{Singh}},\ }\href@noop {} {\bibfield  {journal} {\bibinfo  {journal} {Funct.
				Mater. Lett.}\ }\textbf {\bibinfo {volume} {3}},\ \bibinfo {pages} {223}
		(\bibinfo {year} {2010}{\natexlab{b}})}\BibitemShut {NoStop}%
	\bibitem [{\citenamefont {Spitzer}(1970)}]{spitzer1970lattice}%
	\BibitemOpen
	\bibfield  {author} {\bibinfo {author} {\bibfnamefont {D.}~\bibnamefont
			{Spitzer}},\ }\href@noop {} {\bibfield  {journal} {\bibinfo  {journal} {J.
				Phys. Chem. Solids}\ }\textbf {\bibinfo {volume} {31}},\ \bibinfo {pages}
		{19} (\bibinfo {year} {1970})}\BibitemShut {NoStop}%
\end{thebibliography}
	
	%% else use the following coding to input the bibitems directly in the
	%% TeX file.
	
	% \begin{thebibliography}{00}
		
%apsrev4-2.bst 2019-01-14 (MD) hand-edited version of apsrev4-1.bst
%Control: key (0)
%Control: author (8) initials jnrlst
%Control: editor formatted (1) identically to author
%Control: production of article title (0) allowed
%Control: page (0) single
%Control: year (1) truncated
%Control: production of eprint (0) enabled
%

		% \end{thebibliography}
\end{document}